\documentclass[12pt,number]{elsarticle}
\usepackage[latin9]{inputenc}
\usepackage{float}
\usepackage{hyperref}
\hypersetup{pdftitle={Li-ion battery paper},
	    	pdfauthor={Ata Mesgarnejad, Northeastern University},
	    	colorlinks,
	    	pdfcreator={pdflatex},
	    	unicode=false,
	    	pdftoolbar=false,
	    	pdfmenubar=true,
	   		pdffitwindow=true,
	   		pdfnewwindow=true,
	    	linkcolor=red,
	    	citecolor=red,
	    	filecolor=black,
	    	urlcolor=blue,
	    	}
\usepackage{mathrsfs}
\usepackage{amsthm}
\usepackage{amsmath}
\usepackage{amssymb}
\usepackage{graphicx}
\usepackage{pbox}
\usepackage{esint}
\usepackage{nomencl}
\usepackage{diagbox,ragged2e}
\usepackage{algpseudocode}
\makeatletter
\usepackage[normalem]{ulem}

\floatstyle{ruled}

\newfloat{algorithm}{tbp}{loa}
\floatname{algorithm}{Algorithm}
\theoremstyle{plain}

\theoremstyle{definition}

\theoremstyle{remark}

\usepackage{amsxtra}
\usepackage{mathrsfs}
\usepackage{amsthm}
\usepackage{pdfsync}
\usepackage[version=4]{mhchem}

\usepackage{xargs}                      
\usepackage[pdftex,dvipsnames]{xcolor}

\newcommand{\Tr}{\mathrm{Tr}}

\newcommand{\C}{\mathcal{C}}
\newcommand{\Cr}{\mathcal{C}_r}

\newcommand{\J}{\mathcal{J}}

\newcommand{\ie}{\emph{i.e., }}
\newcommand{\eg}{\emph{e.g., }}

\newcommand{\etal}{\emph{et al.}}

\newcommand{\El}{\mathrm{E}}

\newcommand{\micron}{\mu\mathrm{m}}
\newcommand{\R}{\mathcal{R}}
\newcommand{\cm}{c_{\max}}
\newcommand{\F}{\mathcal{F}}
\newcommand{\Fs}{\mathscr{F}}
\newcommand{\Jh}{\hat{J}}

\makeatother
  \providecommand{\definitionname}{Definition}
  \providecommand{\remarkname}{Remark}

\begin{document}
\let\today\relax

\title{Phase Field Modeling of Chemomechanical Fracture of Intercalation Electrodes:\\
Role of Charging Rate and Dimensionality}
\author[AM]{A.~Mesgarnejad}


\author[AK]{A.~Karma\corref{cor1}}
\cortext[cor1]{Corresponding author}

\ead{a.karma@northeastern.edu}
\address{$^1$ Center for Inter-disciplinary Research on Complex Systems, Department of Physics, Northeastern University, Boston, MA. 02115, U.S.A.}

\begin{abstract}
 
We investigate the fracture of Li-ion battery cathodic particles using a thermodynamically consistent phase-field approach that can describe arbitrarily complex crack paths and captures the full coupling between Li-ion diffusion, stress, and fracture. Building on earlier studies that introduced the concept of electrochemical shock, we use this approach to quantify the relationships between stable or unstable crack propagation, flaw size, and C-rate for 2D disks and 3D spherical particles. We find that over an intermediate range of flaw sizes, the critical flaw size for the onset of crack propagation depends on charging rate as an approximate power-law that we derive analytically. This scaling law is quantified in 2D by exhaustive simulations and is also supported by 3D simulations. In addition, our results reveal a significant difference between 2D and 3D geometries. In 2D, cracks propagate deep inside the particle in a rectilinear manner while in 3D they propagate peripherally on the surface and bifurcate into daughter cracks, thereby limiting inward penetration and giving rise to complex crack geometries.

\end{abstract}

\maketitle

\section{Introduction}\label{sec:introduction}

With the demand for electric vehicles and hand-held electronics on the rise, research on rechargeable batteries and, specifically lithium-ion batteries, becomes increasingly important. 
The need to understand the failure mechanism of these batteries is essential for increasing their life span. 
Chemo-mechanical failure is one of the primary modes of degradation. 
The fracture of cathodic and anodic particles due to intercalation-induced stresses has been extensively studied experimentally \cite{Gabrisch:2008a,Wang:2012a,Wang:2013}. 
The creation of new fracture surfaces impairs the performance of the batteries due to the loss of electrical contact \cite{Chakraborty:2013,Zavalis:2013} and the creation of solid electrolyte interfaces~(SEI) that promotes the irreversible loss of lithium (Li) ions~\cite{Peled:1979,Deshpande:2012}.

On the theoretical side, problems arising from the interplay of diffusion and mechanics have been long considered in the literature.
Prussin~\cite{Prussin:1961} and Lawrence~\cite{Lawrence:1966} were among the first to study the creation and motion of dislocations due to diffusion induced misfit strains.
Subsequently, Liu~\etal~\cite{Liu:1970} studied the attraction of corrosive solutes to the crack tip.
In the context of chemo-mechanical fracture, Huggins and Nix~\cite{Huggins:2000} studied the initiation of fracture due to the intercalation-driven misfit stresses using a simple 1D model of a thin film bilayer. 
In this bilayer geometry with a rigid substrate, a constant misfit strain caused by Li intercalation in the thin film is sufficient to create cracks by a mechanism similar to thermal expansion~\cite{Leon-Baldelli:2014}.
Therefore, using a classical Griffith criterion~\cite{Lawn:1993},  Huggins and Nix were able to derive a critical film thickness for fracture.
The extension to free-standing particles was subsequently considered in several studies~\cite{Christensen:2006,Christensen:2010,Bhandakkar:2010,Woodford:2010}. Unlike in a thin film constrained on a substrate, a uniform concentration does not create stresses in a free-standing particle. However, due to the finite time to diffusively homogenize the Li concentration inside a particle,
a concentration gradient that produces stresses can nonetheless be created when the Li ion flux through the particle boundary, \ie the charging rate (C-rate), is sufficiently large. By analogy with thermal shock, Woodford et al.~\cite{Woodford:2010,Woodford:2012,Woodford:2013} coined the term ``electrochemical shock'' to describe this mode of C-rate dependent fracture ~\cite{Bourdin:2014a}.   
A consistent model that considered the effect of intercalation induced stresses on diffusion was used by Christensen~\etal~\cite{Christensen:2006,Christensen:2010} to obtain a failure criterion solely based on the magnitude of the resulting stresses.
Bhandakkar and Gao~\cite{Bhandakkar:2010} investigated the initiation of a periodic array of equidistant cracks in a thin strip under an imposed constant galvanostatic flux.
Using a cohesive zone model and neglecting the effect of fracture on the concentration field, they derived a scaling law relating the largest ``safe'' strip thickness, below which cracks do not initiate, and C-rate. 
Woodford~\etal~\cite{Woodford:2010} investigated the propagation of fracture from an initial radial penny-shaped flaw on a spherical particle.
Their calculation of the stress intensity factors at the crack tip was carried out in a simplified geometry, also neglecting the effect of fracture on the concentration field. Their results show that, at a given C-rate, both continuous and abrupt propagation modes are possible for different initial flaw sizes. They further show that for a given flux, there exists a largest safe particle size that does not fracture for any flaw size.
Their numerical results indicate that this critical particle size scales as a power-law of charging rate (C-rate).

Even though those studies have yielded quantitative predictions of the dependence of safe particle size on C-rate~\cite{Bhandakkar:2010,Woodford:2010}, they do not consider the full coupling between elasticity, fracture, and diffusion. In addition, penny-shaped cracks are assumed to remain coplanar as they penetrate a 3D spherical particle \cite{Woodford:2010}. In practice, more complex non-coplanar crack patterns may develop that depend on C-rate. 
The goal of this article is to investigate the fracture of 
Li-ion battery cathodic particles using a thermodynamically consistent phase-field approach that captures the full coupling between elasticity, fracture, and diffusion, and that can describe arbitrarily complex crack paths. We exploit those advantages to quantify the relationship between crack propagation, flaw size, and C-rate, and to describe for the first time complex 3D crack patterns.  
Due to their variational formulation, phase-field models of fracture~\cite{Francfort:1998,Karma:2001a,Bourdin:2008a}, offer a unique methodology to tackle the simulation of chemo-mechanical crack growth.
These models have been validated by theoretical analyses \cite{Hakim:2009} and comparisons of predicted and observed crack paths in non-trivial geometries~\cite{Mesgarnejad:2015}. 
They have been used to reproduce complex experimental observations in brittle fracture including thin-film fracture~\cite{Mesgarnejad:2013}, thermal fracture \cite{Bourdin:2014a}, mixed mode fracture~\cite{Chen:2015b}, dynamic fracture~\cite{Chen:2015b,Chen:2017a,Lubomirsky:2018}, fracture in colloidal systems~\cite{Peco:2019}, ductile fracture~\cite{Mozaffari:2015,Ambati:2015,Borden:2016,Alessi:2017}, and fatigue fracture~\cite{Alessi:2018a,Carrara:2018,Mesgarnejad:2018}.
Given their potential over the past few years, researchers have extended the use of these models to chemo-mechanical fracture in battery particles~\cite{Miehe:2015,Zuo:2015,Klinsmann:2016,Klinsmann:2016a}.
This approach has already been used~\cite{Klinsmann:2016,Klinsmann:2016a} to corroborate findings of Woodford~\etal~\cite{Woodford:2010} such as the existence of unstable and stable 
crack propagation as a function of initial flaw size and to describe simple 3D crack patterns. 

In this article, we extend and generalize the results of previous studies~\cite{Bhandakkar:2010,Woodford:2010,Klinsmann:2016} to account for the effect of the crack length on the failure of 2D circular disks and 3D spherical particles.
By performing an exhaustive series of 2D simulations for different flaw sizes and C-rates for a fixed particle radius, we identify three regimes of fracture propagation where (I) large flaws comparable to the particle size do not propagate due to insufficient driving stresses, (II) for intermediate flaw sizes, the critical flaw size scales as an approximate power-law function of C-rate with an exponent that we derive analytically, (III) for very small flaw sizes the C-rate required for propagation diverges resulting in a flux-independent minimum flaw size.
Next, we obtain a scaling law relating the safe particle size, computed with a fixed flaw size to particle radius ratio, to the C-rate.
Furthermore, we show that the topology of fracture changes profoundly in 3D compared to 2D.
We find that, unlike in our 2D studies and previous 3D studies~\cite{Woodford:2010,Klinsmann:2016}, where coplanar cracks penetrate radially towards the center of the particle, 3D cracks remain mostly superficial and branch to tile complex crack patterns on the particle surface.
Our results indicate that, despite this difference, the dependence of critical flaw size on C-rate follows a similar power-law scaling as in 2D. 


This article is organized as follows. In Section~\ref{sec:formulation}, we outline a thermodynamically consistent formulation of chemo-mechanical concentration and stress evolution and fracture.
In Section~\ref{sec:numerical-implementation}, we carry out a scaling analysis of the governing equations and define a subset of key dimensionless parameters. In Section~\ref{sub:chemo-mechanical-fracture-flux}, for a generic set of material parameters for \ce{LiMn2O4}, we present the results of an extensive set of numerical simulations in 2D circular disks and examine the propagation of radial flaws of different sizes.
We investigate the influence of C-rate and initial flaw size on crack stability, generalizing the findings of~\cite{Woodford:2010,Klinsmann:2016}.
We extend our analysis to maximal C-rates in \ref{sub:chemo-mechanical-fracture-dirichlet} and show that there exists a safe particle size regardless of the initial flaw size that can be predicted based on material properties including fracture energy, elastic modulus, and magnitude of misfit strain.
We finally extend our analysis to 3D spherical particles with a single radial penny-shaped surface flaw in Section~\ref{sub:3D-Calculations}.
Lastly, in Section~\ref{sec:conclusion}, we summarize our main findings and point out possible future extensions. 

\section{Formulation}\label{sec:formulation}
We define the total free energy $\Fs(u,c,\Gamma)$ for a domain $\Omega\subset\mathbb{R}^{n}$, containing the crack set $\Gamma\subset\Omega$, for displacement $u$ and concentration $c$

\begin{equation}\label{eq:total-energy}
	\Fs(u,c,\Gamma)= \Fs_{el}(u,c,\Gamma)+\Fs_{c}(c)+\Fs_{\Gamma}(\Gamma)
\end{equation}
where $\Fs_{el}$ is the elastic energy, $\Fs_{\Gamma}(\Gamma)$ is the energetic cost of fracture and $\Fs_{c}$ is the free energy due to the intercalation of Li ions. We can write the elastic energy as
\begin{equation}\label{eq:elastic-energy1}
	\Fs_{el}(u,c,\Gamma)= \int_{\Omega\setminus\Gamma} \mathcal{W}(u,c)\,dx
\end{equation}
where we define the elastic strain energy as $\mathcal{W}(u,c)=\sigma_{ij}\varepsilon_{ij}/2$ in which, the elastic strain is defined as $\varepsilon_{ij}(u,c)=e_{ij}(u)-\epsilon_{0}(c-c_0)\delta_{ij}$ where $\epsilon_{0}$ is the volume expansion coefficient. 
Furthermore, we define the linear strain $e_{ij}(u)=(u_{i,j}+u_{j,i})/2$, and the Cauchy stress tensor $\sigma_{ij}(u,c)=\C_{ijkl}\varepsilon_{kl}$.
Moreover, for Lame's constants $\lambda,\mu$ the isotropic elasticity tensor is written as $\C_{ijkl}=\lambda\delta_{ij}\delta_{kl}+\mu(\delta_{ik}\delta_{jl}+\delta_{il}\delta_{jk})$.

We write the free energy of the Li ions intercalating in the host lattice as~\cite{McKinnon:1983}
\begin{equation}\label{eq:diffusion-energy}
	\Fs_{c}(c)= \int_{\Omega} f_{c}(c)\,dx
\end{equation}
where
\begin{equation}\label{eq:solute-energy}
	f_{c}=\cm\R T\left[\frac{c}{\cm}\ln\left(\frac{c}{\cm}\right)+\left(1-\frac{c}{\cm}\right)\ln\left(1-\frac{c}{\cm}\right)\right]
\end{equation}
is the entropy of mixing for an ideal binary solution, $\cm$ is the maximum concentration achievable when all accommodating sites are filled, $\R$ is the gas constant, and $T$ is the absolute temperature.

In the spirit of brittle fracture, we write the energetic cost of creating fracture surfaces as
\begin{equation}\label{eq:surface-sharp}
	\Fs_{\Gamma}(\Gamma)= G_c\mathcal{H}^{n-1}(\Gamma)
\end{equation}
where $G_{c}$ is the energy required to create a unit area (unit length in 2D) of new cracks, $\mathcal{H}^{m}$ is the $m$--dimensional Hausdorff measure (\ie $\mathcal{H}^{2}(\Gamma)$ is the aggregate area and $\mathcal{H}^{1}(\Gamma)$ is the aggregate length of cracks $\Gamma$ in three and two dimensions, respectively).

\subsection{Phase-field model}\label{sub:phase-field-model}

We use the phase-field model to approximate the sharp interface free energy~\eqref{eq:total-energy} by introducing a fracture phase field $\phi$ and an associated length-scale $\xi$. 
Roughly speaking, as $\xi\to0$,  the displacement field minimizing~\eqref{eq:ATE} converges to that of minimizing~\eqref{eq:total-energy}, the field $\phi$ converges to 1 almost everywhere and goes to zero near the cracks.
In this article, we treat the length-scale $\xi$ as a regularization parameter to study the sharp-interface limit of the phase-field model that reduces to classical linear elastic fracture mechanics \cite{Hakim:2009}.
We write the approximate free energy replacing $\Fs_{el}(u,c,\Gamma)$ by $\Fs_{el}(u,c,\phi)$ and $\Fs_{\Gamma}(\Gamma)$ by $\Fs_{\phi}(\phi)$ as
\begin{equation}\label{eq:ATE}
	\Fs(u,c,\phi)= \Fs_{el}(u,c,\phi)+\Fs_{c}(c)+\Fs_{\phi}(\phi)
\end{equation}
with the elastic energy
\begin{equation}\label{eq:elastic-energy}
	\Fs_{el}(u,c,\phi)= \int_{\Omega} g(\phi)\mathcal{W}(u,c)\,dx
\end{equation}
and the energetic cost of fracture  
\begin{equation}\label{eq:surface-pf}
	\Fs_{\phi}(\phi)= \frac{G_c}{4C_{\phi}}\int_{\Omega}\left(\frac{w(\phi)}{\xi}+\xi \left|\nabla \phi \right|^2\right)\,dx
\end{equation}
where $C_{\phi}=\int_{0}^{1}\sqrt{w(\phi)}\,d\phi$ is a scaling constant. 
In the past decade, there has been a growing trend in studying a broad class of \emph{rate independent gradient damage models} in the form of~\eqref{eq:ATE} \cite{Pham:2012,Pham:2013}. 
In this article, we use the Karma-Kessler-Levine model~(\textbf{KKL})~\cite{Karma:2001a,Hakim:2009} defined using $g(\phi)=4\phi^3-3\phi^4$, $w(\phi)=1-g(\phi)$. 
This model allows us to follow the propagation of a fracture from a single flaw by prohibiting the initiation of new cracks in undamaged material (\ie $\phi=1$).


\subsection{Diffusion equation}\label{sub:diffusion}
Following the classical argument of continuity (mass conservation), we write the diffusion equation for concentration as~\cite{McKinnon:1983}
\begin{equation}\label{eq:mass-conservation}
	\frac{\partial c}{\partial t} = -\nabla\cdot J
\end{equation}
where $J$ is the flux of Li ions. 
We define the flux as the product of the mobility and the gradient of chemical potential
\begin{equation}\label{eq:flux}
	J=-a(\phi)M(c) \nabla\mu
\end{equation}
where the mobility of Li ions in the host lattice $M(c)=m_0\,(c/\cm)(1-c/\cm)$ first increases and then decreases as a function of relative concentration $c/\cm$. 
We define the chemical potential as
\begin{equation}
	\mu=\frac{\delta \Fs}{\delta c}
\end{equation}
in which $\dfrac{\delta \Fs}{\delta c}$ is the Fr\'echet derivative of free energy $F$ with respect to the concentration $c$ and can be written as
\begin{equation}\label{eq:dFdc}
	\frac{\delta \Fs}{\delta c}= \frac{df_c}{dc}(c) -\epsilon_{0}g(\phi)\sigma_{kk}
\end{equation}
replacing from above in~\eqref{eq:flux} we finally get:
\begin{align}
	\frac{\partial c}{\partial t}&=\nabla\cdot \left[m_0 a(\phi)\left[\R T\nabla c + \left(\frac{c}{\cm}\right)\left(1-\frac{c}{\cm}\right)\nabla \psi\right]\right]\label{eq:diffusion}\\
	\psi&=-\epsilon_{0}\cm\,g(\phi)\sigma_{kk} \label{eq:axuilliary-diffusion}
\end{align}
We note that \eqref{eq:diffusion} represents a Fickian diffusion with the second term coupling to the mechanical hydrostatic stresses.

As a first estimate, we make the crack surface completely permeable to Li ion diffusion similar to~\cite{Klinsmann:2016} by introducing $a(\phi)=1$. 
We motivate this choice by noticing that the electrolyte will leak inside the newly created fracture surfaces thus making them permeable to ion transfer.
The precise choice of $a(\phi)$ should be determined by further study of specific material and the interaction of the electrolyte and fracture surfaces and is out of the scope of this article.

Furthermore, to model galvanostatic charging, we write the flow of Lithium ions from the boundary as a given imposed flux $\Jh$:
\begin{equation}
	\left.J\right|_{\partial_f\Omega}=\frac{i}{\F}\equiv\Jh \label{eq:flux-BC}
\end{equation}
where $i$ is the surface current density and $\F$ is Faraday's constant.

The galvanostatic boundary condition of form~\eqref{eq:flux-BC} is a first order approximation of ion transfer on the particle boundary and. 
A more general study can be done by prescribing the value of the flux on the surface at a given voltage as using the Butler-Volmer equation to model reaction kinetics on the cathodic surface of the particle~\cite{Doyle:1993}.
We should note that, as for any diffusion process of a bounded field (here $0\leq c\leq \cm$) with a flux boundary, this boundary condition cannot be maintained for any arbitrary value of $\Jh$ for an infinite time. 
In particular, $\Jh\gg1$ at $t\ll R^2/m_0\R T$ a depleted boundary layer is created where the concentration at the boundary reaches zero and the flux cannot be maintained any longer. 

\section{Numerical implementation}\label{sec:numerical-implementation}

\subsection{Dimensional analysis}\label{sub:nD}

For the flux boundary condition~\eqref{eq:flux-BC}, it is pertinent to introduce the nominal charging time as the time required to fill the volume of the particle $V$ with a surface flux $\Jh$ acting on surface area $A$ \ie
\begin{equation}\label{eq:t_C}
	t_C=\frac{\cm V}{\Jh A}
\end{equation}
It is also customary in Li-ion literature to introduce the so-called charging rate $C_r=1/t_C$, which is usually measured in $\mathrm{hr}^{-1}$ units.
To perform the numerical simulations, we adimensionalize the spatial dimensions by the particle radius $R$, the concentration by $\cm$, the time by the diffusion time $t_D=R^2/D_0$ where $D_0=m_0\R T$ is the diffusion constant, and the stresses by energy per unit volume $\cm\R T$.
We write dimensionless charging rate $\Cr$ as
\begin{equation}\label{eq:Cr}
	\Cr=t_D C_r=\frac{t_D}{t_C}
\end{equation}
and the dimensionless flux as
\begin{equation}\label{eq:BC-flux-ND}
	\J=\Cr\frac{\bar{V}}{\bar{A}}=\frac{t_D}{t_C}\frac{\bar{V}}{\bar{A}}=\frac{\Jh R}{\cm D_0}
\end{equation}
where $\bar{V}=V/R^n$ is the dimensionless volume of the particle, and $\bar{A}=A/R^{n-1}$ is the dimensionless surface area of the flux boundary.
The the dimensionless charging rate $\Cr$ can also be understood intuitively as a mechanical loading parameter noticing that the driving force for crack propagation is the gradient of the concentration field in~\eqref{eq:elastic-energy} that is controlled by the flux $\Jh$.
As a result, for low dimensionless charging rates $\Cr<1$ (where the nominal charging time is long compared to the diffusion time $t_C\gg t_D$) the concentration will homogenize and thus creates no misfit stresses.

We should also highlight the important dimensionless numbers that uniquely define the simulations performed, namely the relative strength of the elastic energy compared to the chemical energy $\El/\cm\R T$, Poisson's ratio $\nu$, maximum misfit strain $\beta=\cm\epsilon_{0}$, and the relative domain geometry \ie radius and initial flaw size, compared to the Griffith length scale $R/(G_c/\El)=R/l_G$ and $a_0/l_G$.

\subsection{Governing equations}\label{sub:governing-eq}
To implement our numerical simulations using the Galerkin finite element method we introduce the weak forms of the governing equations.
The governing equations for the concentration diffusion is derived from its flow rule~\eqref{eq:diffusion}-\eqref{eq:axuilliary-diffusion} by multiplying both sides with test functions and integrating by parts. We also incorporate an implicit time integration scheme to ensure the accuracy and stability of the integration.
\begin{align}
	&\int_{\Omega}\left(\frac{c_{t}-c_{t-1}}{\delta t}\right)\,\tilde{c}\,dx+\int_{\Omega}a(\phi_{t-1}) \nabla_{\Theta} c\cdot \nabla\tilde{c}\,dx\nonumber\\
	&\quad\quad\quad+\int_{\Omega}a(\phi_{t-1})\,M(c_{t})\nabla_{\Theta}\psi\cdot \nabla\tilde{c}\,dx+\int_{\partial_{f}\Omega} \J\tilde{c}\,ds=0~&\forall\tilde{c}\in H^1(\Omega)\label{eq:gov-c}\\
	&\int_{\Omega} \left[\psi_{t}+\beta g(\phi_{t-1})\sigma_{kk}(u_{t-1},c_{t})\right]\tilde{\psi}\,dx=0~&\forall\tilde{\psi}\in H^1(\Omega)\label{eq:gov-axuilliary-c}
\end{align}
where $n$ is the surface normal to $\partial_f\Omega$, subscripts denote time steps with $\delta t$ as the time step size, and we define $\nabla_{\Theta}\{\circ\}=(1-\Theta)\nabla\,\{\circ\}_t+\Theta\nabla\,\{\circ\}_{t-1}$ as the implicit gradient operator associated with time-fraction $\Theta$. 
In all calculations in this paper we used $\Theta=0.5$ which corresponds to the midpoint method and results in a second order accurate and unconditionally stable time integration for concentration field $c$.

Moreover, since in practical systems the time-scale of elasticity and fracture propagation are orders of magnitude smaller than that of diffusion, we assume that they are instantaneous.
In this setting, we seek the minimizers for the displacement field $u$ and the fracture phase-field $\phi$ for each time step $t_i$. 
Hence, the governing equations for displacement (\ie elasticity) and fracture phase-field, are written as Euler-Lagrange equations of the total energy~\eqref{eq:total-energy} for displacement field $u$ and phase field $\phi$

\begin{align}
	&\int_{\Omega} g(\phi_t)\,\sigma_{ij}(u,c) e_{ij}(\tilde{u})\,dx=0~&\forall \tilde{u}\in H^1(\Omega)\label{eq:gov-elasticity}\\
	&\int_{\Omega} \left[\frac{dg}{d\phi}(\phi_t)\,\mathcal{W}(u,c)\right]\,\tilde{\phi}\,dx\nonumber\\&+\frac{G_c}{4C_{\phi}}\int_{\Omega}\left[\frac{1}{\xi}\left(\frac{dw}{d\phi}(\phi_t)\right)\tilde{\phi}+2\xi\nabla\phi\cdot\nabla\tilde{\phi}\right]\,dx=0~&\forall\tilde{\phi}\in H^1(\Omega)\label{eq:gov-phi}
\end{align}
where $c_t$ is the concentration given by the solution of~\eqref{eq:gov-c}-\eqref{eq:gov-axuilliary-c}.

\subsection{The solution algorithm}\label{sub:solution-algorithm}
The phase-field fracture method requires that the spatial resolution of discretization to resolve the characteristic approximation length $\xi$. 
The resulting problems are often very large and necessitate the use of a parallel programming paradigm and the complex numerical tools therein. 
Our implementation relies on the distributed data structures provided by \texttt{libMesh}~\cite{libmesh} and for linear algebra on \texttt{PETSc}~\cite{petsc-efficient,petsc-user-ref}.
On the other hand, we assume that elasticity and fracture are instantaneous and write their governing equations as the weak forms of Euler-Lagrange equation for minimizers of~\eqref{eq:ATE} with respect to displacement field $u$ and phase field $\phi$ respectively (see~\ref{sub:governing-eq} for details).
This is roughly equivalent to the limit of vanishing relaxation time $\tau_{\phi}\rightarrow0$ assuming that the phase field $\phi$ follows Ginsburg-Landau gradient dynamics:
\begin{equation}
	\tau_{\phi}\frac{\partial \phi}{\partial t}=\frac{1}{\cm\R T}\frac{\delta {\Fs}}{\delta \phi}
\end{equation}

We use a classical alternate minimization algorithm~\ref{algo:alt-min}~\cite{Bourdin:2008a} since the governing equations for elasticity and phase-field are only convex in either $u$ or $\phi$ when the other is kept constant~\cite{Bourdin:2008a}.
It is also worth mentioning that to enforce irreversibility of fracture and ensure boundedness of phase field~$0\leq\phi\leq1$ and the relative concentration~$0\leq c\leq 1$, we use a bounded reduced space Newton minimization scheme for the discrete energy provided in \texttt{PETSc}~\cite{petsc-efficient,petsc-user-ref}.

\begin{algorithm}[!ht]
\begin{algorithmic}[1]
\State{Set $\phi_{0}=1$ in bulk at $\phi=0$ at the initial crack.}
\State{Let $\delta_{altmin}$ be given tolerance parameters.}
\For{$n=0$ to $N$}
	\State{Update $c_{n}$ based on $u_{n-1}$ and $\phi_{n-1}$~\eqref{eq:gov-c}-\eqref{eq:gov-axuilliary-c}}.
	\State{Initialize the phase field from last time step: $\phi^{0}\longleftarrow\phi_{n-1}$}.
	\While{$\left| \phi^j-\phi^{j-1}\right|_{L^\infty} \ge \delta_{altmin}$}
		\State{Update $u^{j+1}$ using $\phi^{j}$ and $c_n$~\eqref{eq:gov-elasticity}}.
		\State{Update $\phi^{j+1}$ using $u^{j+1}$ and $c_n$~\eqref{eq:gov-phi}}.
		\State{$j\longleftarrow j+1$}.
	\EndWhile
	\State{Store the converged time step $u_{n}\longleftarrow u^{j}$ and $\phi_{n}\longleftarrow\phi^{j}$.}
\EndFor
\end{algorithmic}
\caption{The alternate minimization algorithm. Subscripts are time steps while superscripts denote the internal alternate minimization iteration.}
\label{algo:alt-min}
\end{algorithm}

\section{Numerical results}\label{sec:numerical-results}
In the following section, we focus on the numerical simulation of a cathodic particle at the time of charging. 
We first present our two-dimensional results for circular particles with a preexisting radial flaw on its surface under galvanostatic and potentiostatic boundary conditions.
Subsequently, we analyze the fracture of spherical particles with penny-shaped cracks in three dimensions. 

\subsection{Chemo-mechanical fracture of circular particles: (I) galvanostatic (flux) boundary condition}\label{sub:chemo-mechanical-fracture-flux}

The misfit strains generated during charging and discharging processes can lead to the creation and propagation of cracks in Li-ion battery particles. 
For a preexisting flaw on the surface of a cathodic particle, the removal of Li-ions during the charging process causes the outer layer of the particle to contract faster than its inner core.
Therefore, for fast enough charging rates, the region of tensile stresses created in the outer periphery can activate surface defects creating cracks that will then propagate through the particle.
In this section, we present the results of numerical simulation for the fracture of circular particles induced by the removal of Lithium ions during charging.
Our goal is two fold: (i) to understand the activation and propagation of a preexisting flaw in a circular cathodic particle, (ii) 2D simulations also enable us to combine the results of many such simulations to give insight into critical parameters for the design of these particles. 

\begin{figure}[htb!]
	\begin{center}
		\includegraphics[width=.4\columnwidth]{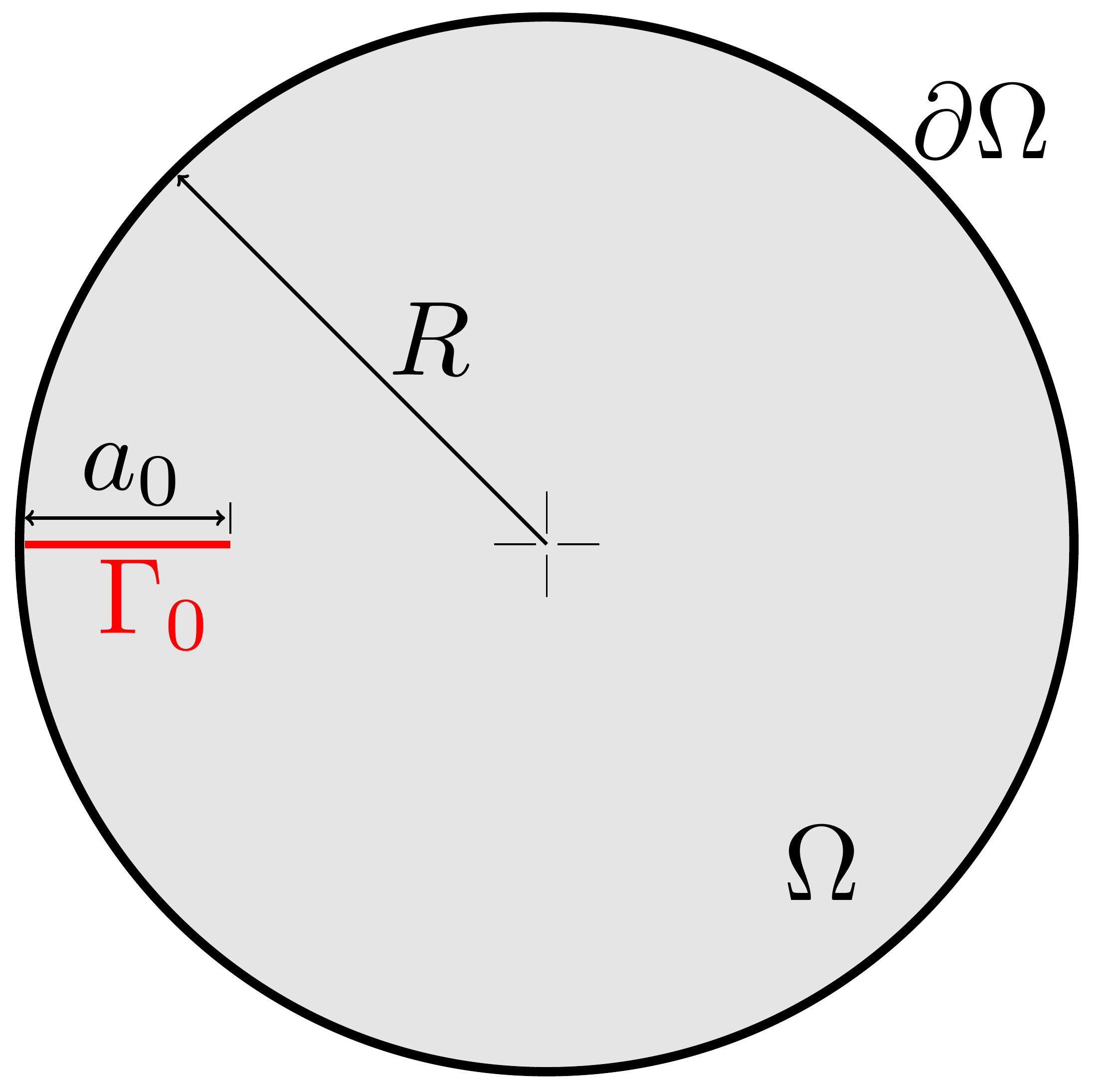}
	\end{center}
	\caption{The schematics of 2D chemo-mechanical fracture of circular particles numerical simulations.}
	\label{fig:schematics}
\end{figure}

Fig.~\ref{fig:schematics} shows a schematics for this problem.
We assume a preexisting radial crack $\Gamma_0$ of length $a_0$ and impose a dimensionless galvanostatic flux~$\J$ (corresponding to the dimensionless charge-rate~$\Cr=t_D/t_C$ according to~\eqref{eq:BC-flux-ND}).
As stated before, we treat phase-field length scale $\xi$ as a regularization of Griffith brittle fracture. Therefore for the Griffith length scale defined as
\begin{equation}\label{eq:Griffith-length}
	l_G=\frac{G_c}{\El}
\end{equation}
we use $\xi>l_G$.
For these numerical simulations we use a constant relative process zone size $\xi/R=1.25\times10^{-2}$ for relative flaw size $a_0/R>0.1$ and use $\xi/a_0=5$ for $a_0/R<0.1$ for optimal use of computational resources. 
Table~\ref{tab:matprop-chemo-mechanical-fracture} summarizes the material properties corresponding to \ce{LiMn2O4} used in our simulations.

\begin{table}[htb!]
\centering
\caption{Material properties of \ce{LiMn2O4} for numerical simulation of chemo-mechanical fracture~\cite{Woodford:2010}.}
\vspace{.5em}
\begin{tabular}{l l l l}
\hline
Property & Symbol & Units & Value \tabularnewline
\hline 
Elastic Modulus & $\El$ & $\mathrm{N}\,\mathrm{m}^{-2}$ & $2\times10^{11}$ \tabularnewline
Poisson Ratio & $\nu$ & - & $0.3$ \tabularnewline
Fracture Toughness & $G_c$ & $\mathrm{N}\,\mathrm{m}^{-1}$ & $100$ \tabularnewline
Diffusivity & $D_0$ & $\mathrm{m}^{2}\,\mathrm{s}^{-1}$ & $2.2\times10^{-13}$ \tabularnewline
Maximum Concentration & $\cm$ & $\mathrm{mol}\,\mathrm{m}^{-3}$ & $2.37\times10^{4}$ \tabularnewline
Misfit Strain Constant & $\epsilon_{0}$ & $\mathrm{m}^{3}\,\mathrm{mol}^{-1}$ & $1.09\times10^{-6}$ \tabularnewline
Density & $\rho$ & $\mathrm{kg}\,\mathrm{m}^{-3}$ & $4.28\times10^{3}$ \tabularnewline
Temperature & $T$ & $\mathrm{K}$ & $3\times10^{2}$ \tabularnewline
\hline
Dimensionless Expansion Coefficient & $\beta$ & $\mathrm{m/m}$ & $0.025$ \tabularnewline
Griffith Length Scale & $l_G$ & $\micron$ & $5\times10^{-4}$ \tabularnewline
\hline 
\end{tabular}
\label{tab:matprop-chemo-mechanical-fracture}
\end{table}

\begin{figure}[htb!]
	\begin{center}
		\includegraphics[width=\columnwidth]{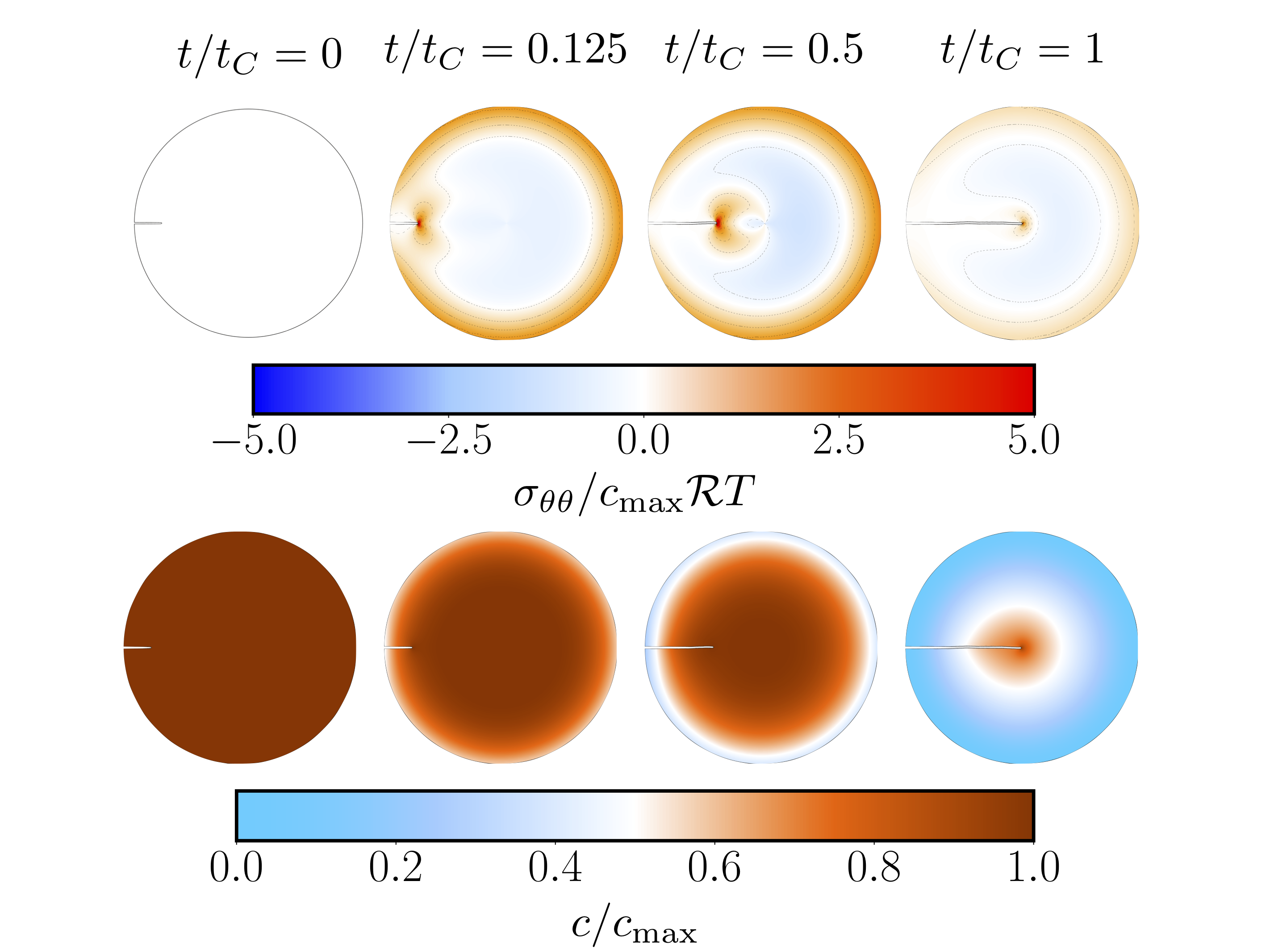}
	\end{center}
	\caption{Time snapshots of crack propagation in chemo-mechanical fracture of a $R/l_G=4.2\times 10^4$ 2D circular particle with a preexisting $a_0/l_G=10^4$ radial crack driven by $\Cr=5.57$ charging rate showing continuous propagation (see also Fig.~\protect\ref{fig:dlc5}). 
	Color codes depict the dimensionless hoop stress $\sigma_{\theta\theta}/\cm\R T$ distribution (top), and the dimensionless concentration distribution $c/\cm$ perturbed as the result of the crack-tip stress field (bottom).}
	\label{fig:R21-a5-crack-propagation}
\end{figure}

\begin{figure}[htb!]
	\begin{center}
		\includegraphics[width=\columnwidth]{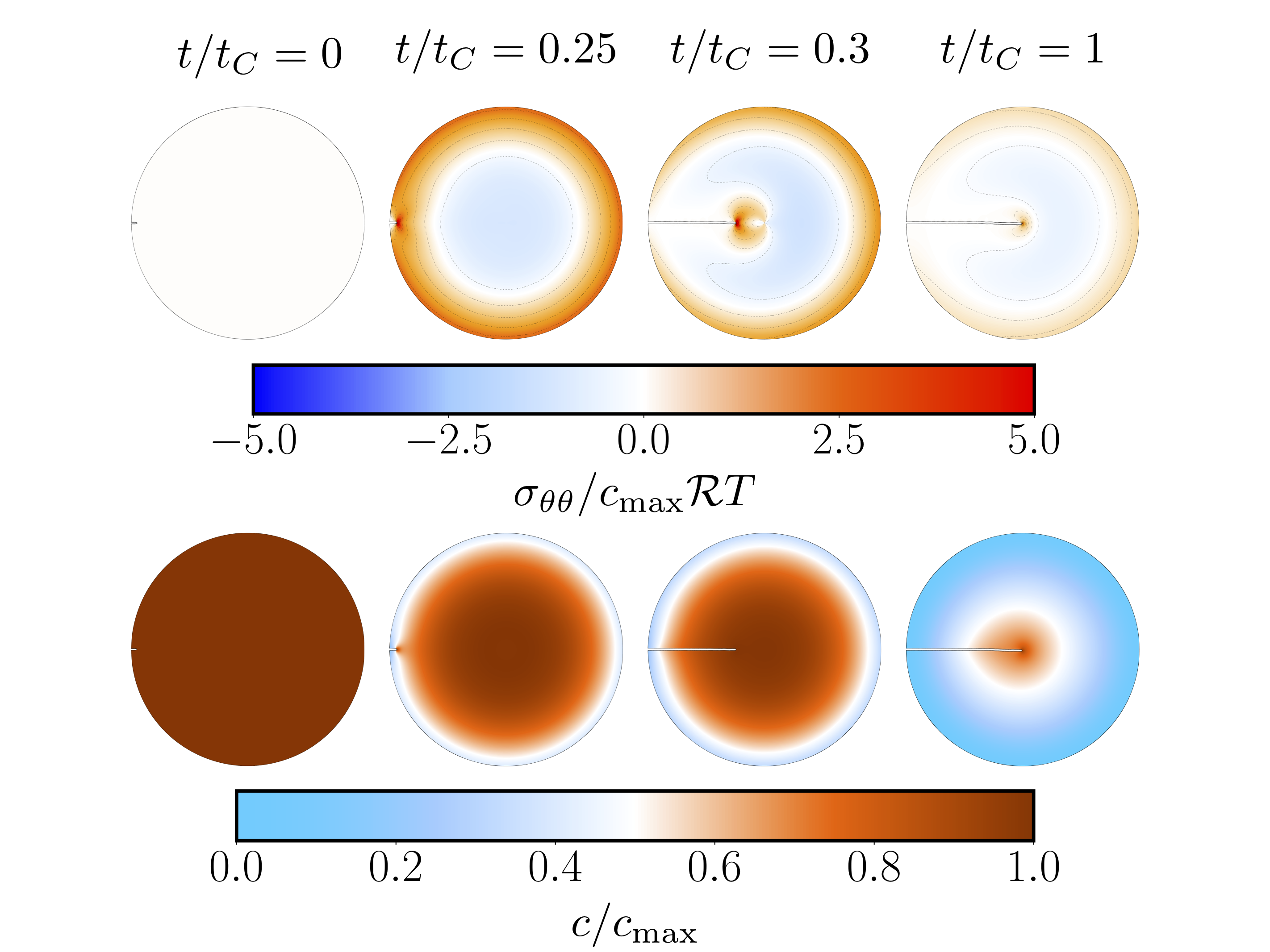}
	\end{center}
	\caption{Time snapshots of crack propagation in chemo-mechanical fracture of a $R/l_G=4.2\times 10^4$ 2D circular particle with a preexisting $a_0/l_G=2\times10^3$ radial crack driven by $\Cr=5.57$ charging rate showing initial $(a-a_0)/R\simeq 0.7$ abrupt propagation for $t/t_C\simeq0.27$ followed by continuous propagation (see also Fig.~\protect\ref{fig:dlc1}). 
	Color codes depict the dimensionless hoop stress $\sigma_{\theta\theta}/\cm\R T$ distribution (top), and the dimensionless concentration distribution $c/\cm$ perturbed as the result of the crack-tip stress field (bottom).}
	\label{fig:R21-a1-Cr5-crack-propagation}
\end{figure}

\begin{figure}[htb!]
	\begin{center}
		\includegraphics[width=\columnwidth]{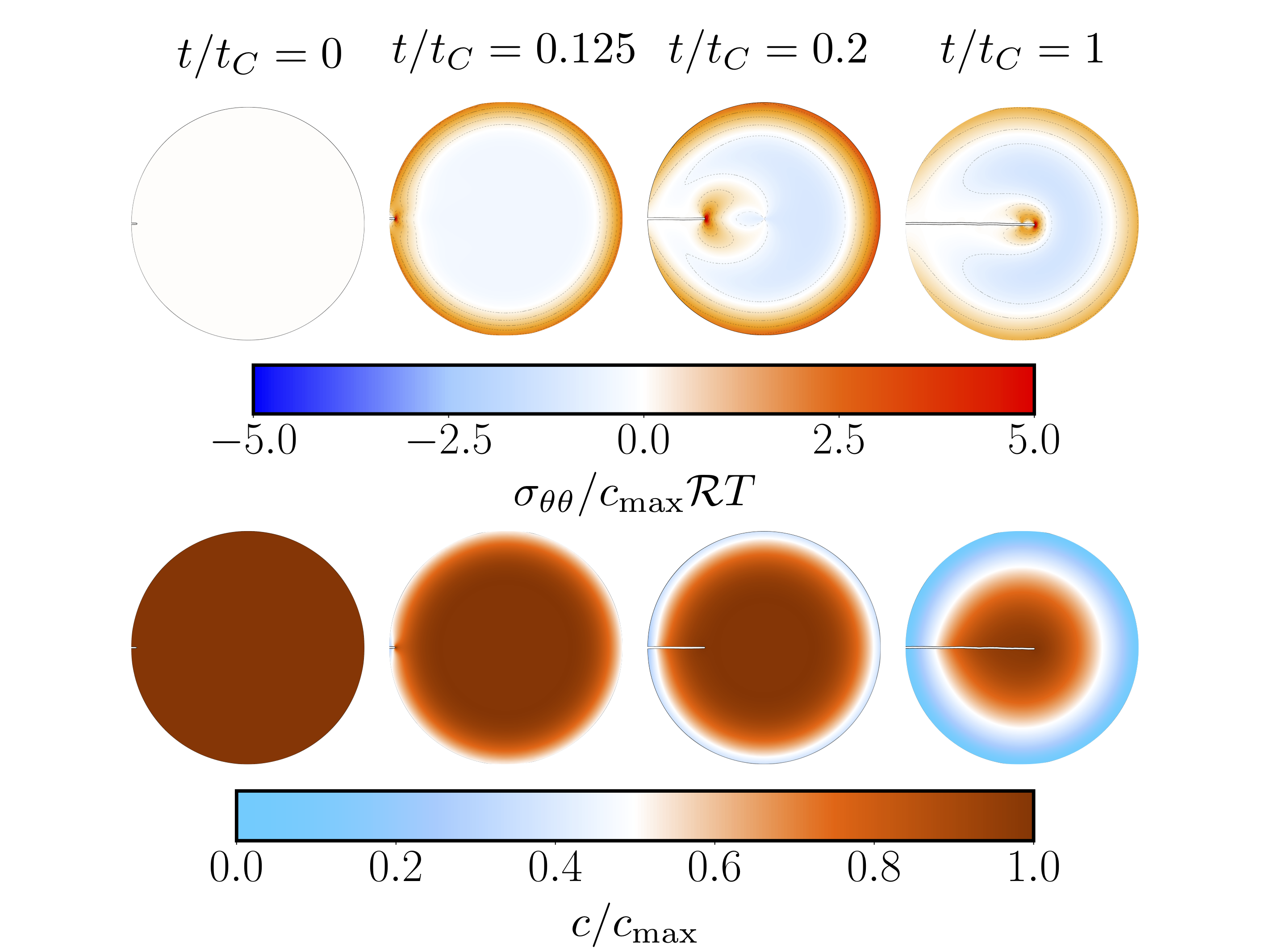}
	\end{center}
	\caption{Time snapshots of crack propagation in chemo-mechanical fracture of a $R/l_G=4.2\times 10^4$ 2D circular particle with a preexisting $a_0/l_G=2\times10^3$ radial crack driven by $\Cr=8.35$ charging rate showing $(a-a_0)/R\simeq 0.3$ abrupt propagation at $t/t_C\simeq 0.13$ followed by continuous propagation (see also Fig.~\protect\ref{fig:dlc1}). 
	Color codes depict the dimensionless hoop stress $\sigma_{\theta\theta}/\cm\R T$ distribution (top), and the dimensionless concentration distribution $c/\cm$ perturbed as the result of the crack-tip stress field (bottom).}
	\label{fig:R21-a1-Cr8-crack-propagation}
\end{figure}

To highlight the mechanism and modes of radial crack penetration in these particles, we first study three sample cases in Figs.~\ref{fig:R21-a5-crack-propagation}--\ref{fig:R21-a1-Cr8-crack-propagation}.
These sample results correspond to fracture of a $R/l_G=4.2\times 10^4$ and $a_0/l_G=2\times10^3,10^4$ initial radial flaws (corresponding to a $R=21\,\micron$ particle with $a_0=1,5\,\micron$ for material properties in Table~\ref{tab:matprop-chemo-mechanical-fracture}) and using $\Cr=5.57,8.35$ dimensionless charging rates (circled gray in Fig.~\ref{fig:min-Cr-flcrack-L21}).
In these simulations, we first compare two cases with different initial flaw lengths at the same dimensionless charging rate $\Cr$ and then study two cases where we keep $a_0$ constant and change the $\Cr$. 
As we stated before, our simulation results show that tensile hoop stresses are created on the periphery of these particles that can then drive the surface flaws to penetrate radially inside the particle (top row in Figs.~\ref{fig:R21-a5-crack-propagation}--\ref{fig:R21-a1-Cr8-crack-propagation}).
Figs.~\ref{fig:R21-a5-crack-propagation} shows that the crack propagation for the larger initial flaw $a_0/l_G=10^4$ under lower dimensionless charging rate $\Cr=5.57$ is continuous.
However, the smaller initial flaw under the same charging rate $\Cr=5.57$ propagates abruptly jumping many process zone sizes (see third columns in Fig.~\ref{fig:R21-a1-Cr5-crack-propagation}).
The abrupt propagation occurs in the context of Griffith fracture where the crack releases more energy as it propagates (\ie for the energy release rate defined as $G=-\partial \Fs_{el}/\partial a$ at a frozen concentration (constant load), the crack is unstable if ${dG}/{da}>0$).
Analogous results on abrupt propagation due to misfit strains are also reported in the context of thermally driven cracks. 
For example, Bahr~\etal~\cite{Bahr:1987} explicitly calculate the energy release rate as a function of crack length for thermal-quenching-induced cracks.

A similar contrast between continuous and abrupt propagation is also predicted by Woodford~\etal~\cite{Woodford:2010} where they explicitly calculate the mode-I stress intensity factor $K_I$ (and, by symmetry since $K_{II}\equiv0$, the energy release rate $G$) for a radial penny-shaped crack in a spherical particle.
Their calculations show that for some choices of particle size and initial flaw size ${dG}/{da}>0$; therefore, for such flaw sizes, propagation is unstable and abrupt.
Concurring with our simulations, their calculations (figure 5. in~\cite{Woodford:2010} where their results correspond directly to Figs.~\ref{fig:R21-a5-crack-propagation}--\ref{fig:R21-a1-Cr5-crack-propagation}) show that the flaws smaller than $a_0/l_G<4\times 10^3$ will propagate abruptly given our choice of parameters and vice versa. 

The transition from abrupt to continuous propagation can also be understood in analogy with mechanical loading in standard fracture mechanics configurations. 
At lower charging rates where the concentration field has to penetrate on the scale of the particle size before the energy release rate reaches the fracture energy, the activation of the initial notch is analogous to a crack in half plane under constant far field opening stress that results in an unstable propagation.
On the other hand, at high charging rates where the concentration field penetrated on the scale of the initial notch only, the problem resembles a compact specimen with the crack opening from the back that results in stable propagation.
Our hypothesis is further verified, comparing Fig.~\ref{fig:R21-a1-Cr5-crack-propagation} to Fig.~\ref{fig:R21-a1-Cr8-crack-propagation}. 
We can see that at the higher $\Cr$ (higher flux), the initial abrupt crack propagation is shorter.
While surprising at first glance, we can explain the longer abrupt fracture propagation at lower $\Cr$, noticing that the flaw is activated earlier for the higher $\Cr$. 
Thus, at the time of the initial jump, the hoop stresses penetrate farther inside the particle for the lower flux providing more elastic energy to be converted to new fracture surfaces.

\begin{figure}[htb!]
	\begin{center}
		\includegraphics[width=\columnwidth]{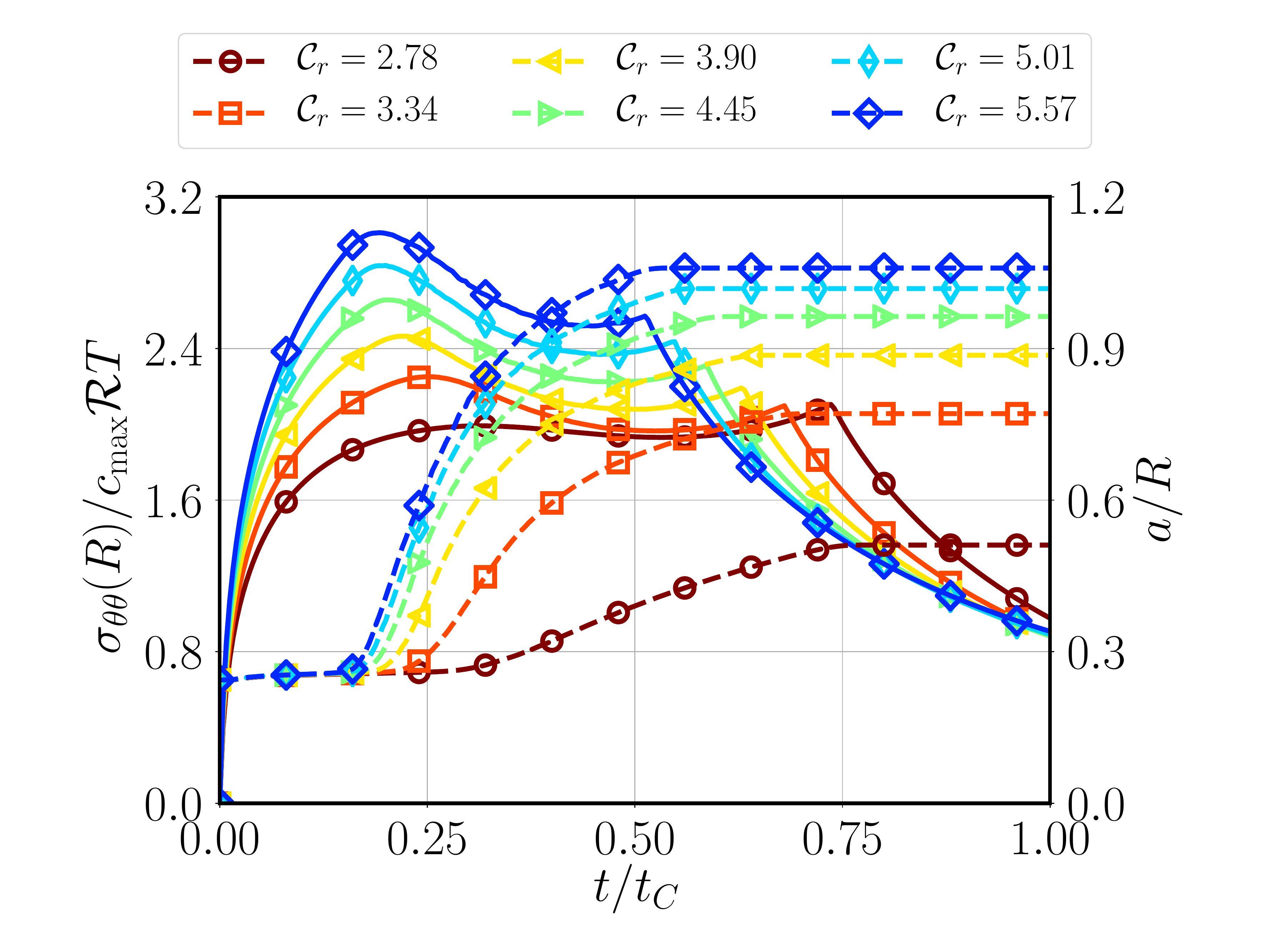}
	\end{center}
	\caption{Numerical results of a $R/l_G=4.2\times 10^4$ particle with an initial $a_0/l_G=10^4$ flaw vs charging time fraction ${t}/{t_C}$ for different dimensionless charging rates $\Cr$: showing relative crack length increase $a/R$ (dashed lines, right vertical axis) and maximum surface hoop stress far from the crack-tip $\sigma_{\theta\theta}(r=R)$ (solid lines, left vertical axis).
	Time snapshots for evolution of $\Cr=5.57$ was previously shown in Fig.~\protect{\ref{fig:R21-a5-crack-propagation}}.
	}
	\label{fig:dlc5}
\end{figure}

\begin{figure}[htb!]
	\begin{center}
		\includegraphics[width=\columnwidth]{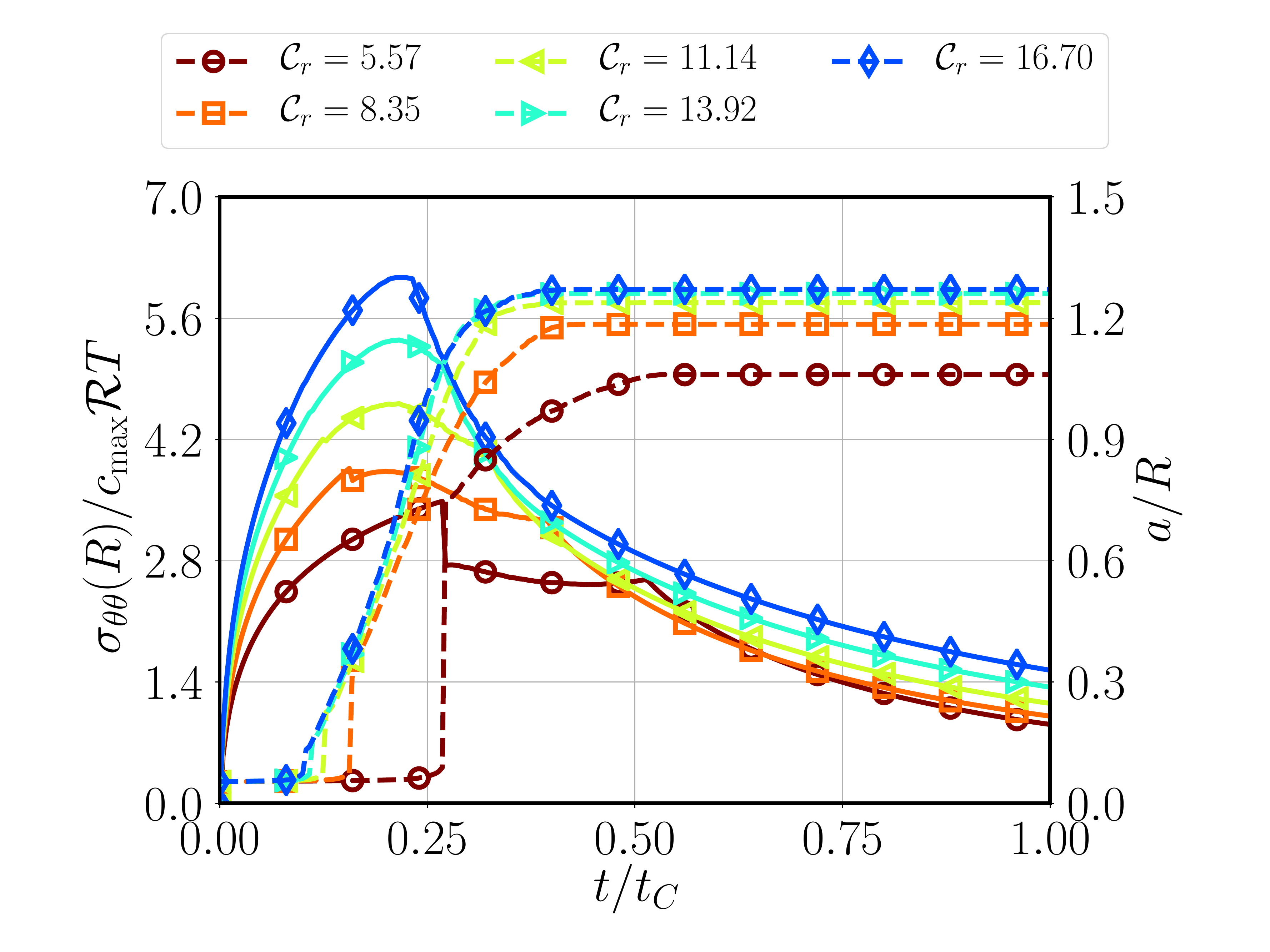}
	\end{center}
	\caption{Numerical results of a $R/l_G=4.2\times 10^4$ particle with an initial $a_0/l_G=2\times 10^{3}$ flaw vs charging time fraction ${t}/{t_C}$ for different dimensionless charging rates $\Cr$: showing relative crack length increase $a/R$ (dashed lines, right vertical axis) and maximum surface hoop stress far from the crack-tip $\sigma_{\theta\theta}(r=R)$ (solid lines, left vertical axis). Time snapshots for evolution of $\Cr=5.57$ and $\Cr=8.35$ were previously shown in Fig.~\protect{\ref{fig:R21-a1-Cr5-crack-propagation}} and Fig.~\protect{\ref{fig:R21-a1-Cr8-crack-propagation}} respectively.
	}
	\label{fig:dlc1}
\end{figure}

\begin{figure}[htb!]
	\begin{center}
		\includegraphics[width=\columnwidth]{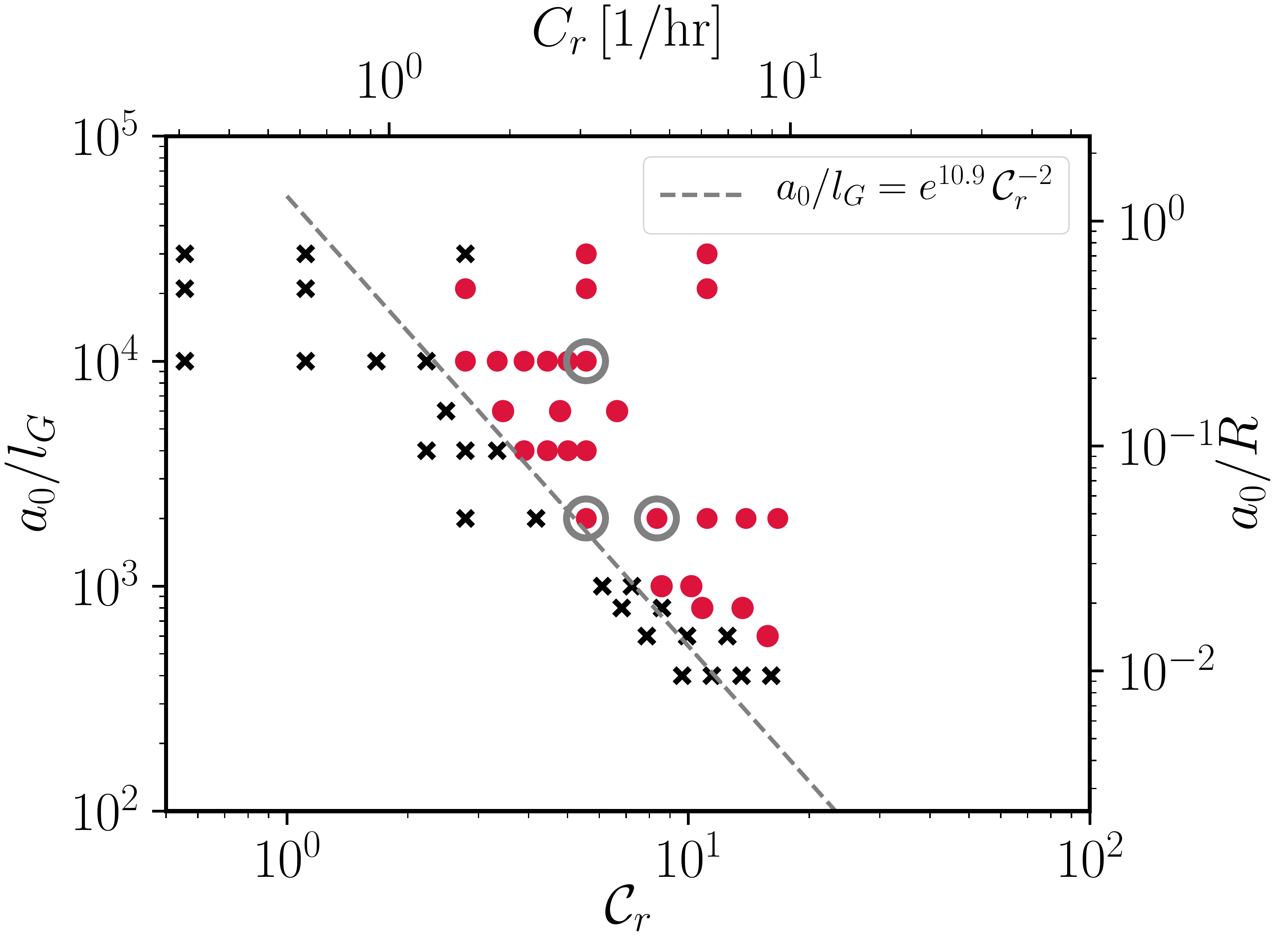}
	\end{center}
	\caption{Numerical simulation results for a $R/l_G=4.2\times 10^4$ particle with $a_0/l_G=200\textup{--}10^4$ initial flaws: flaw activation diagram for the dimensionless charge rate $\Cr$ vs initial flaw size $a_0/l_G$. Circles show the activated vs. crosses show the unactivated cracks. The gray dashed line shows the power-law $a_0/l_G=e^{10.9}\,\Cr^{-2}$ ($a_0\beta^2/l_G=35.93\,\Cr^{-2}$). Computations corresponding to Figs.~\ref{fig:R21-a5-crack-propagation}--\ref{fig:R21-a1-Cr8-crack-propagation} are circled in gray.}
	\label{fig:min-Cr-flcrack-L21}
\end{figure}

Different modes of fracture propagation are further demonstrated and quantified in Figs.~\ref{fig:dlc5}--\ref{fig:dlc1} where we show the evolution of relative crack lengths $a/R$ and the dimensionless hoop stress in front of the crack at $r=R$ versus the charging time fraction ${t}/{t_C}$ for the simulations of a $R/l_G=4.2\times10^4$ particle with $a_0/l_G=2\times10^3,~10^4$ preexisting radial flaws respectively and for different dimensionless charging rates $\Cr$ (including cases highlighted in Figs.~\ref{fig:R21-a1-Cr5-crack-propagation}--\ref{fig:R21-a1-Cr8-crack-propagation}).
Similar to Fig.~\ref{fig:R21-a5-crack-propagation}, the simulations for the larger $a_0/l_G=10^4$, depicted in Fig.~\ref{fig:dlc5}, show a clear trend whereby the crack is activated $t/t_C\simeq0.2$ and propagates smoothly.
Unlike results presented in Figs.~\ref{fig:dlc5}, the initial crack propagation in Fig.~\ref{fig:dlc1} is abrupt and decreases with higher $\Cr$.  
We should also note, in Figs.~\ref{fig:dlc5}--\ref{fig:dlc1}, that while the abrupt initial propagation is bigger for the smaller flux, the cracks extend farther into the particle for higher fluxes in line with our intuitive understanding that higher fluxes provide more energy.
Similar observations were also made in~\cite{Klinsmann:2016} (see figure 15 in the reference) where for a $R/l_G=4.65\times10^{10}$ particle containing $a_0/l_G=9.3\times10^8$ initial crack they observed a larger initial abrupt propagation for lower fluxes and vice versa.

Our self-consistent simulations also allow us to observe the interaction of the stresses with the concentration field.
We note that the tensile crack-tip stresses attract ions from its vicinity and results in crack tip enrichment.
Many semi-analytic simulations, currently available in the literature, are based on the radial approximation of the concentration field~\cite{Bhandakkar:2010,Woodford:2010} (\ie $c=c(r)$) which is calculated for an un-cracked particle.
We study the crack-tip enrichment further in the~\ref{app:crack-tip-enrichment} where in a simple setting we identify an enrichment length scale where its ratio to Griffith length scale, scales as the ratio of maximum misfit stresses to the chemical energy squared (see~\eqref{eq:rc}).



\begin{figure}[htb!]
	\begin{center}
		\includegraphics[width=\columnwidth]{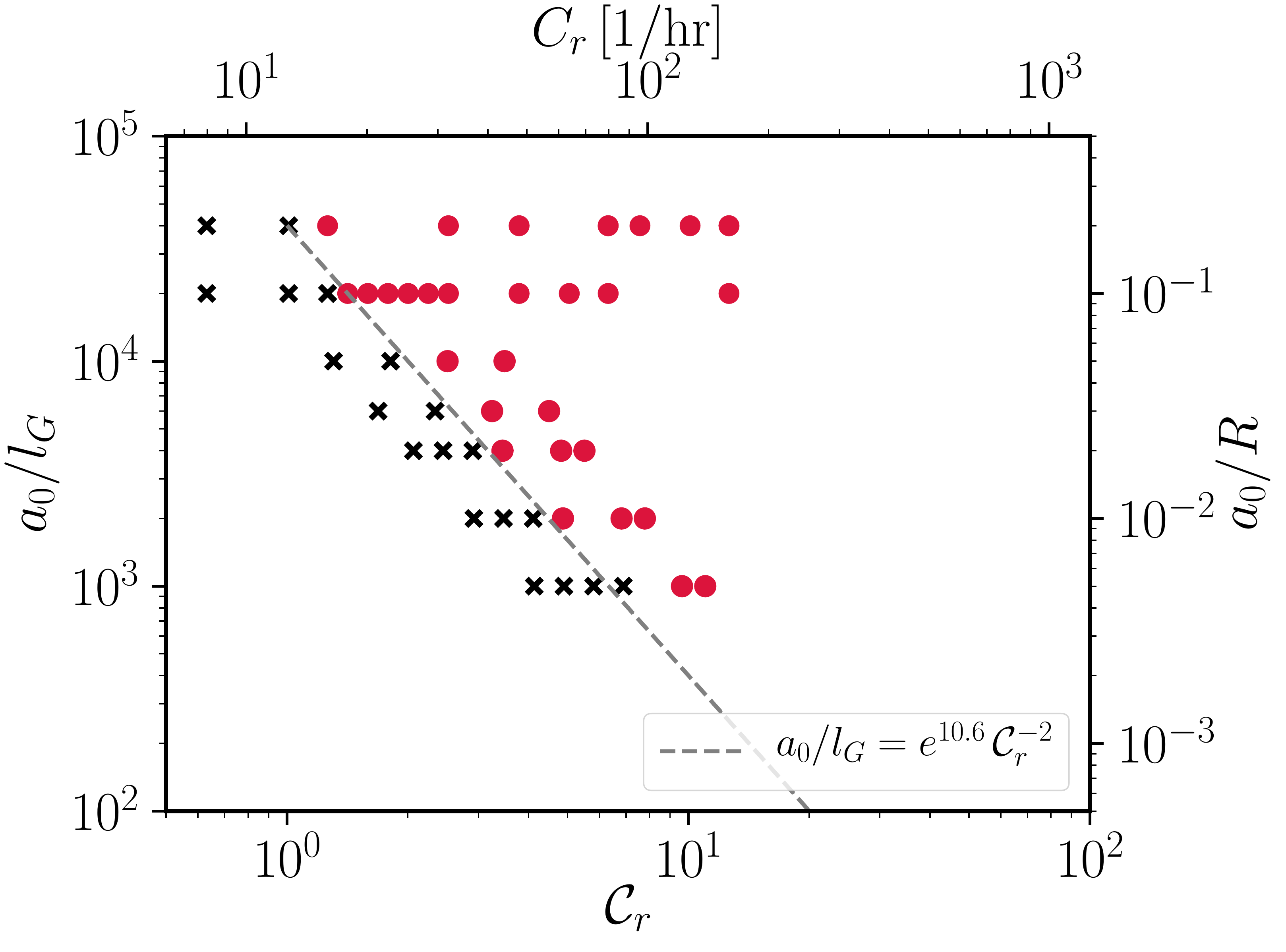}
	\end{center}
	\caption{Numerical simulation results for a $R/l_G=2\times 10^5$ particle with $a_0/l_G=10^3\textup{--}4\times10^4$ initial flaws: flaw activation diagram for the dimensionless charge rate $\Cr$ vs initial flaw size $a_0/l_G$. Circles show the activated vs. crosses show the unactivated cracks. The gray dashed line shows the power-law $a_0/l_G=e^{10.6}\,\Cr^{-2}$ ($a_0\beta^2/l_G=26.62\,\Cr^{-2}$).}
	\label{fig:min-Cr-flcrack-L100}
\end{figure}

With the propagation mechanism elucidated, we now turn our attention to obtaining design parameters for these particles.
Fig.~\ref{fig:min-Cr-flcrack-L21} shows a combined diagram for results of our simulations for $R/l_G=4.2\times 10^4$ particles containing initial flaws of different sizes.
In this phase diagram, the circles mark activated cracks and cross marks depict those not activated at a given charging rate $\Cr$. 
As expected, longer radial cracks need lower charging rate to activate, but very long initial flaws do not propagate since the hoop stresses around the crack tip never grow large enough.
We duplicated the simulations for a larger $R/l_G=2\times 10^5$ particle in Fig.~\ref{fig:min-Cr-flcrack-L100}. 

We can make three important observations from Figs.~\ref{fig:min-Cr-flcrack-L21}--\ref{fig:min-Cr-flcrack-L100};
First the simulations show that there exists a safe charging rate $\mathcal{C}_{r,\min}\simeq2,1$ for $r/l_G=4.2\times10^4,2\times10^5$ particles respectively where flaws regardless of their size do not propagate.
Secondly, for moderate dimensionless charging rates $\Cr=O(1)$ ($t_D\sim t_C$), the minimum flaw size activated $a_{0,\min}$ decreases as the inverse of the dimensionless charging rate squared \ie $a_{0,\min} \sim \Cr^{-2}$.
Thirdly, in Figs.~\ref{fig:min-Cr-flcrack-L21}--\ref{fig:min-Cr-flcrack-L100}, there exists a minimum flaw size $a_{0,\min}^{\infty}$ that flaws smaller, regardless of the dimensionless charging rate do not propagate.

The inverse square law can be understood by noting that the at moderate fluxes the concentration needs to penetrate at the particle length scale before there is enough energy for the flaw to activate (see Figs.~\ref{fig:R21-a5-crack-propagation}--\ref{fig:R21-a1-Cr8-crack-propagation}).
It is worth noting that the propagation at these critical fluxes is always abrupt only for smaller initial notches as detailed previously in this section (see Fig.~\ref{fig:dlc1}).
Therefore the time to activate an initial flaw is similar to the diffusion time of ions in particle size $t_0\sim t_D=R^2/D$. 
Using the mass conservation, we can write that the mass accumulated in the particle is equal to the mass of the ions inserted through its surface \ie $ R^2\Delta c\sim R\hat{J}\,t_0 \sim R^3\hat{J} /D$. 
Rearranging the terms, we find the characteristic variation of Li ion's concentration across the particle scales as $\Delta c\sim \hat{J}R/D$. 
This variation generates maximum hoop stress $\sigma_{\theta\theta}\sim \El\beta \Delta c/\cm$ at the particle surface.
According to the standard Griffith criterion, this stress can activate a flaw of size $a_{0,\min}\sim G_c\El/\sigma_{\theta\theta}^2$. 
Combing the above expressions for $\sigma_{\theta\theta}$, $\Delta c$, $t_D$, $t_C$, and using~\eqref{eq:Griffith-length} we obtain the prediction
\begin{equation}
a_{0,\min}\sim{l_G}(\beta\Cr)^{-2}. \label{eq:inverse-sq-law}
\end{equation}
In other words, the ratio of flaw size to the Griffith length-scale $a_{0,\min}/l_G$, scales as the inverse square of dimensionless charging rate times the maximum misfit strain \ie $a_{0,\min}/l_G=A(\beta\Cr)^{-2}$ where $A$ is a scaling constant.
We can now identify the ``misfit length scale''
\begin{equation}\label{eq:lc}
	l_c=\frac{l_G}{\beta^2}
\end{equation}
which takes into account that the magnitude of maximum misfit stresses generated $\El\beta$ is the appropriate measure of stresses in diffusion driven fracture.
We should highlight that similar length scale was also used in \cite{Bourdin:2014a} for the study of thermally driven cracks.

Using our phase-field simulations presented in Figs.~\ref{fig:min-Cr-flcrack-L21}--\ref{fig:min-Cr-flcrack-L100} we can identify the scaling constants for the two particle sizes as $A=35.93, 26.62$ for $R/l_G=4.2\times10^4,2\times10^5$ particles respectively.
Perhaps not surprisingly, since equation~\eqref{eq:inverse-sq-law} has many simplifications and does not encode all particle size dependencies.
Most notably it ignores the effect of the relative initial flaw size compared to the particle radius $a_0/R$, where changing the relative size of the initial flaw will change the evolution of the concentration field for the moderate charging rates considered.
Furthermore,  \eqref{eq:inverse-sq-law} also ignores the effects of the crack tip enrichment.
As alluded to before, tensile stresses at the tip attract concentration, this introduces another length scale (see~\eqref{eq:rc}) into the system that, in principle, can introduce dependency on the particle radius.
Therefore, the scaling constant in the case of two different particle sizes are different. 
With the scaling constants extracted from the phase-field simulations we can carry the analysis further and obtain the maximum safe charging rate for a given particle size.
For these practical charging rates, we can rewrite the maximum safe $C_{r,\max}$ below in which no flaws can be activated in a particle of radius $R$ as
\begin{equation}
	C_{r,\max} = \frac{1}{t_{c,\min}}=\frac{D_0}{R^2}\left(\frac{Al_c}{a_{0,\min}}\right)^{1/2}
\end{equation}
This scaling law predicts the most conservative charging rate (minimum charging time $t_C$) in terms of basic material properties.

\begin{figure}[htb!]
	\begin{center}
		\includegraphics[width=\columnwidth]{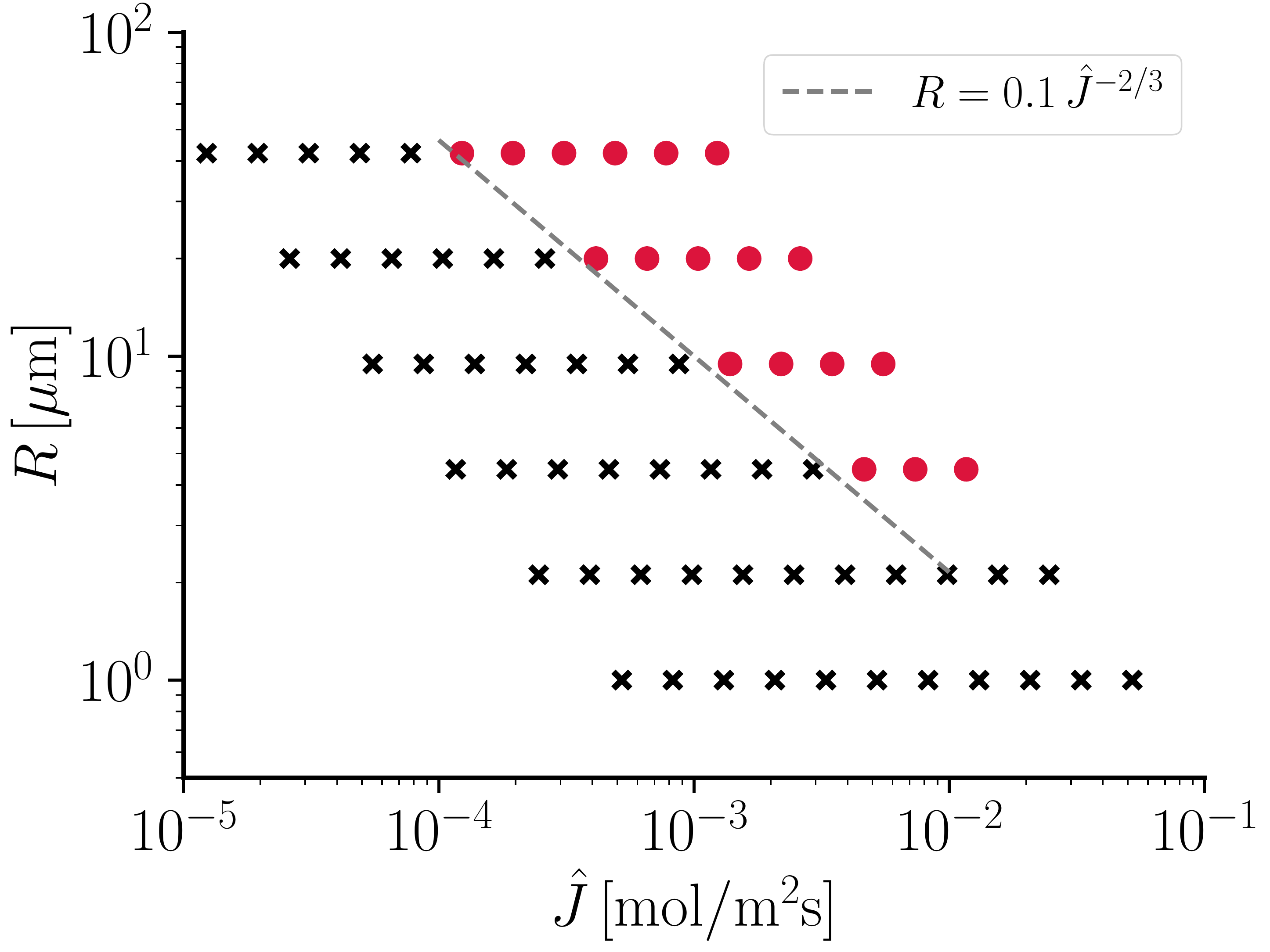}
	\end{center}
	\caption{Flaw activation diagram for circular particles with $a_0/R=0.1$ initial flaw. Circles show the activated vs. crosses show the unactivated cracks. The gray dashed line shows the scaling law $R=0.1\,\Jh^{-2/3}$.}
	\label{fig:R-J}
\end{figure}

Furthermore, to demonstrate the particle size dependency, it is easy to rearrange the power-law in equation~\eqref{eq:inverse-sq-law} for a given dimensionless flaw size $\bar{a}_0=a_0/R$ as $R\sim\Jh^{-2/3}$.
Fig.~\ref{fig:R-J} depicts such a power-law emerging from the combined results of a series of simulations for particles of different size with a $a_0/R=0.1$ initial radial flaw on their surface. 
A similar power-law was independently derived by Bhandakkar and Gao~\cite{Bhandakkar:2010} for initiation of a periodic array of cracks in a thin film using a cohesive zone model. 
There, authors study the initiation of an array of equidistant cracks such that the maximum stress in the system is equal to the cohesive strength of the material under study.
They then, given the fracture energy of the material, investigate whether the displacement opening for the potentially initiated cracks will exceed the critical displacement required to maintain them.
They find that regardless of the cohesive strength of the material, there exists a critical film thickness $H_c\sim \Jh^{-2/3}$ below which no fracture is initiated in the thin film.

Following our third observation, we see that the minimum flaw size activated $a_{0,\min}$ for small initial flaws deviates from the scaling law~\eqref{eq:inverse-sq-law} and approaches a constant value $a_{0,\min}^{\infty}\simeq10^3\,l_G$ for both particle sizes.
At very large $\Cr$ required to activate these minimal flaws, the concentration reaches its minimum physically allowed value $c=0$ at the particle surface in a time $t_0\ll t_D$. 
In the next Section~\ref{sub:chemo-mechanical-fracture-dirichlet}, we study the limit $\Cr\rightarrow \infty$ where $t_0/t_D\rightarrow 0$ by imposing the potentiostatic (Dirichlet) boundary condition $c=0$ at the particle surface. 

\subsection{Chemo-mechanical fracture of circular particles: (II) potentiostatic (Dirichlet) boundary condition}\label{sub:chemo-mechanical-fracture-dirichlet} 

As discussed before for large fluxes $\Cr\rightarrow\infty$, the concentration field reaches its minimum $c=0$ at time $t_0\ll t_C$ as a result of which a depleted boundary layer is created on the surface of the particle. 
Therefore, it is more convenient to study this limit using Dirichlet boundary conditions. 
In this section, we present the results of numerical simulation for fracture of circular particles using $c(R)=0,~t\in[0,t_{\max}]$ Dirichlet boundary condition that is analogous to the maximum flux attainable for this system.
Using this boundary condition, we can find the minimum flaw size activated for different particle radii. 
Similar to the previous section, we chose phase field length scale $\xi$ such that the initial flaw is well resolved (\ie $a_0/\xi\geq4$).

\begin{figure}[htb!]
	\begin{center}
		\includegraphics[width=0.8\columnwidth]{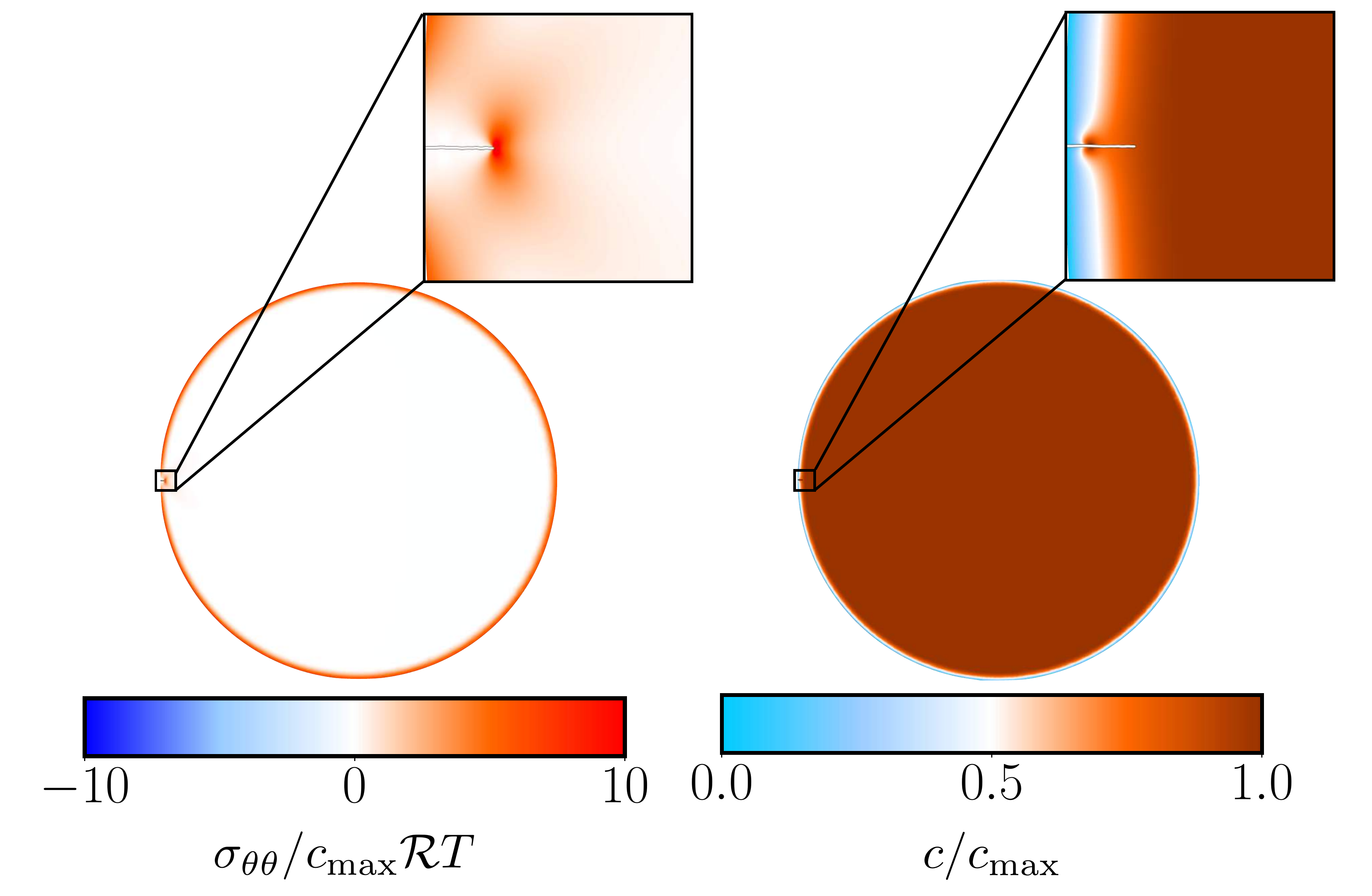}
	\end{center}
	\caption{Numerical simulation results for a $R/l_G=10^5$ particle with $a_0/l_G=400$ initial flaw at first time step after its activation $t/t_D=0.011$ showing the concentration field penetrating at the scale of the initial flaw.
	Color codes depict the dimensionless hoop stress $\sigma_{\theta\theta}/\cm\R T$ distribution (left), and the dimensionless concentration distribution $c/\cm$ perturbed as the result of the crack-tip stress field (right) with inlays showing area near the initial flaw magnified.}
	\label{fig:dirichlet-activation}
\end{figure}

Fig.~\ref{fig:dirichlet-activation} shows the activation of a $a_0/l_G=400$ radial flaw in a $R/l_G=10^5$ particle under Dirichlet boundary conditions.
Unlike the simulations analyzed in the previous section, the initial flaw is activated at $t_a\ll t_D$ in these simulations.
We observe that the fracture propagation stems from the creation of an ion-depleted boundary layer of thickness $h\sim \sqrt{D_0 t_a}$ with a size comparable to the minimum flaw size $a_0$ but much smaller than the particle radius \ie $h\ll R$. 

\begin{figure}[htb!]
	\begin{center}
		\includegraphics[width=\columnwidth]{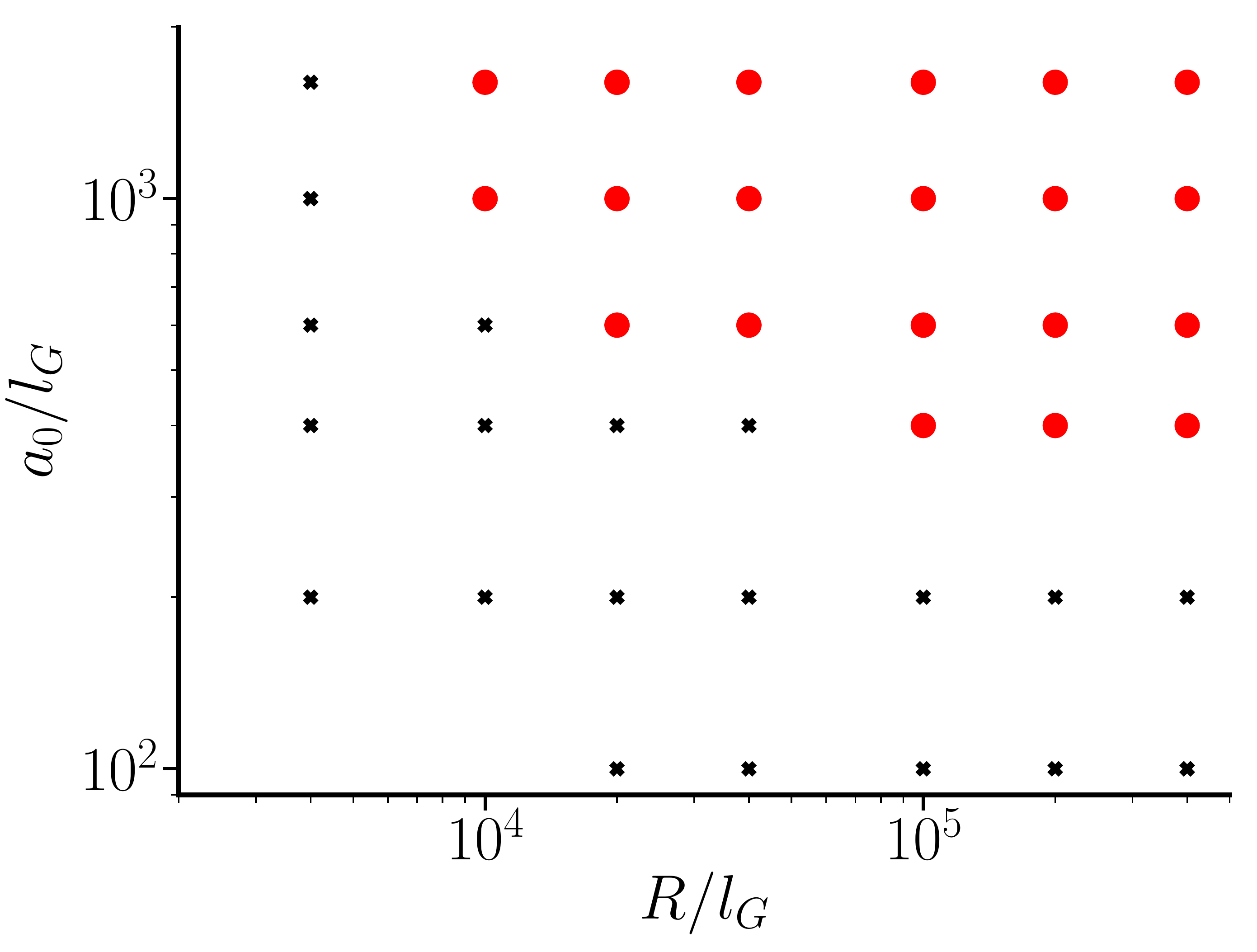}
	\end{center}
	\caption{Flaw activation diagram for circular particles with Dirichlet boundary conditions: results for the dimensionless particle size $R/l_G=2\times10^3\textup{--}4\times10^5$ vs the dimensionless initial flaw size $a_0/l_G=10^2\textup{--}1.6\times10^3$. Circles show the activated vs. crosses show the unactivated cracks.}
	\label{fig:phase-diagram-dirichlet}
\end{figure}

Fig.~\ref{fig:phase-diagram-dirichlet} shows the combined results of these numerical simulations for six particle sizes with different initial flaw sizes where we can make two observations.
Firstly, our numerical results show that for our choice of parameters, there exists a maximum safe particle size $R_{\max}\simeq10^4l_G$ that no flaw would propagate in it.
Simply put, since the minimum activated flaw size decreases with the particle size, it becomes comparable to the particle size for small particles which cannot produce high enough deriving forces to propagate them.
This size is analogous to critical thickness derived in~\cite{Huggins:2000} for a simple 1D bilayer.
Secondly, we notice that as a function of growing particle radius $R\rightarrow\infty$, the smallest flaw activated asymptotically approaches a constant value $a_{0,\min}^{\infty}/l_G\rightarrow5\times10^4$.
Thus the smallest flaw activated for a large particle becomes independent of its radius $R$.
We can elucidate this observation noticing that in the absence of cracks, the maximum hoop stress generated under Dirichlet boundary conditions $\sigma\sim \El\beta$ is independent of the particle radius. 
Therefore, analogous to a flaw on the boundary of a half-space, the flaw size scales $a_{0,\min}^{\infty}\sim G_c\El/(\sigma)^2= l_c$ where the dimensionless prefactor is a function of the particle geometry.
Also, we can relate our observations of minimum flaw size $a_{0,\min}$ at a given radius to results presented in the previous section~\ref{sub:chemo-mechanical-fracture-flux} for large fluxes $\Cr\rightarrow \infty$.
For example, for a $R/l_G=4.2\times10^4$ particle presented in Fig.~\ref{fig:min-Cr-flcrack-L21} the minimum flaw activated is predicted to be $a_0/l_G\simeq4\times10^2$ consistent with the results in Fig.~\ref{fig:phase-diagram-dirichlet}.

\subsection{Fracture of spherical cathodic particles with penny-shaped radial flaws}\label{sub:3D-Calculations}

Although insight gained from the two-dimensional numerical simulations in the previous section is invaluable, only true 3D calculations can hope to capture all essential aspects of chemo-mechanical fracture in these particles.
In this section, we demonstrate the similarities and differences between the simplified 2D and more realistic 3D simulations.
Following the previous section, we model a spherical $R/l_G=2\times10^4$ particle with a penny-shaped flaw on its surface (see~\ref{fig:3D-R10-init-topology}). 
Unlike many similar calculations in 3D (\eg~\cite{Klinsmann:2016a,Klinsmann:2016}), in this article, we simulate the complete sphere without explicit use of any symmetries. 
To this end, the elastic null-space (\ie translation and rotational modes) was calculated and removed prior to the elastic sub-iteration in the alternate minimization algorithm (\ie solving~\eqref{eq:gov-elasticity}). 
Since the computational cost of a uniform fine mesh was prohibitive, we chose a static adaptive meshing scheme, where for $r/l_G>1.6\times10^4$ a fine mesh with an average edge length of $200l_G$ was generated and gradually coarsened to a coarse mesh with an average edge length of $500l_G$ for $r/l_G<1.2\times10^4$. 
This meshing scheme, for different initial flaw sizes, resulted in the computational domain discretized into $\sim13\textup{--}15\,\mathrm{M}$ tetrahedral elements (resulting overall in roughly the same number of degrees of freedom).
Following the 2D simulations, we set the phase-field length scale to $\xi/R=3\times10^{-2}$ and use the material properties as presented in table~\ref{tab:matprop-chemo-mechanical-fracture}.
Moreover, due to the prohibitive cost of the 3D simulations, we limit our investigation to three flaw sizes $a_0/l_G=10^3,1.5\times10^3,2\times10^3$ and charging rates $15\leq \Cr\leq 45$.

\begin{figure}[htb!]
	\begin{center}{}
		\includegraphics[width=\columnwidth]{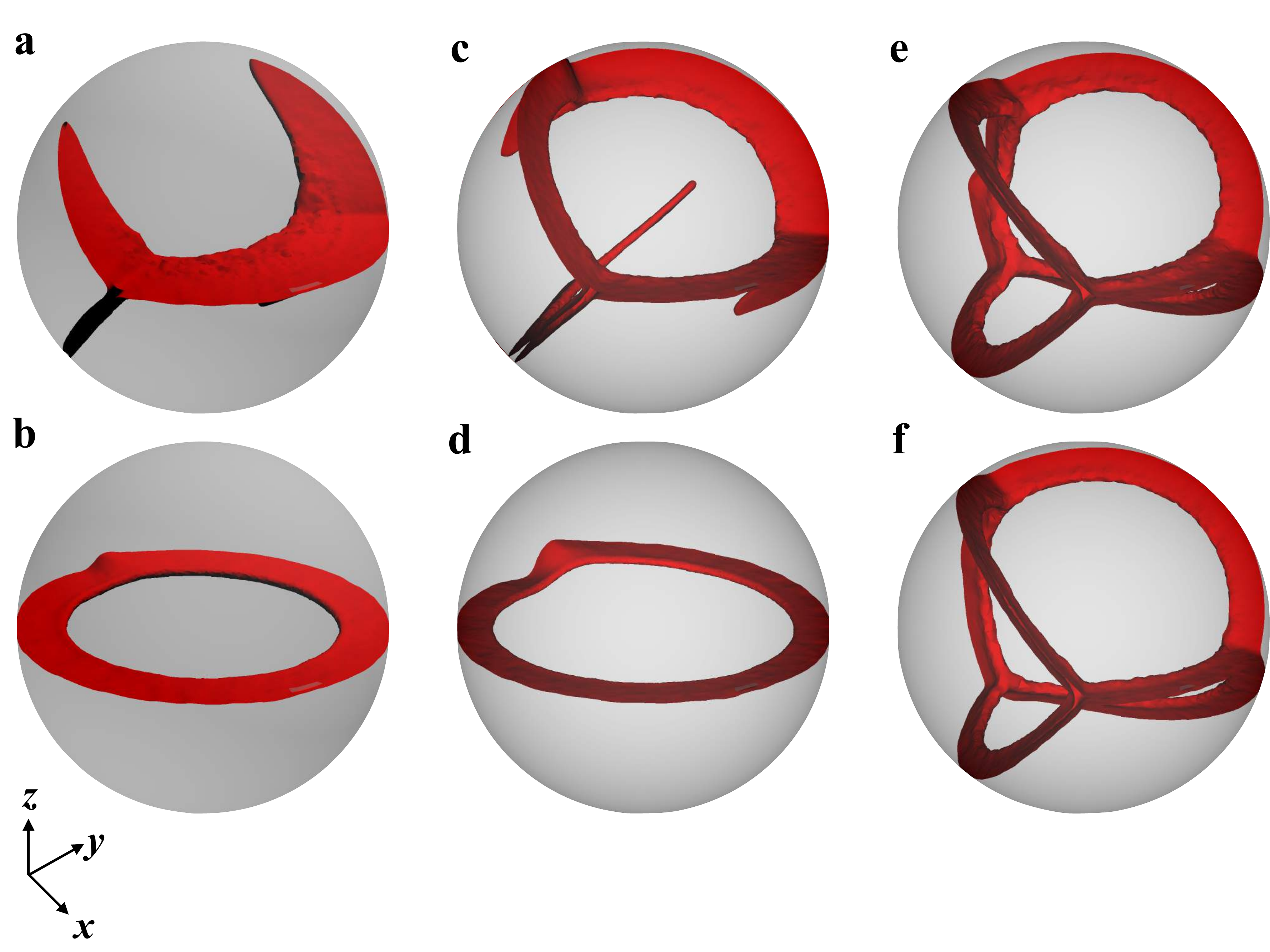}
	\end{center}
	\caption{Numerical simulation results for a $R/l_G=2\times10^4$ particle showing the fracture topology (iso-surface visualization for $\phi=0.5$) after initial abrupt activation (see also Fig.~\protect\ref{fig:3D-phase-diagram}). 
	\textbf{(a)} $a_0/l_G=2\times10^3$ radial penny-shaped crack under $\Cr=21$ at $t/t_C=0.5$.
	\textbf{(b)} $a_0/l_G=2\times10^3$ radial penny-shaped crack under $\Cr=30$ at $t/t_C=0.28$.
	\textbf{(c)} $a_0/l_G=1.5\times10^3$ radial penny-shaped crack under $\Cr=30$ at $t/t_C=0.33$.
	\textbf{(d)} $a_0/l_G=1.5\times10^3$ radial penny-shaped crack under $\Cr=45$ at $t/t_C=0.22$.
	\textbf{(e)} $a_0/l_G=10^3$ radial penny-shaped crack under $\Cr=36$ at $t/t_C=0.32$.
	\textbf{(f)} $a_0/l_G=10^3$ radial penny-shaped crack under $\Cr=45$ at $t/t_C=0.26$.
	}
	\label{fig:topology-3D}
\end{figure}

\begin{figure}[htb!]
	\begin{center}
		\includegraphics[width=.8\columnwidth]{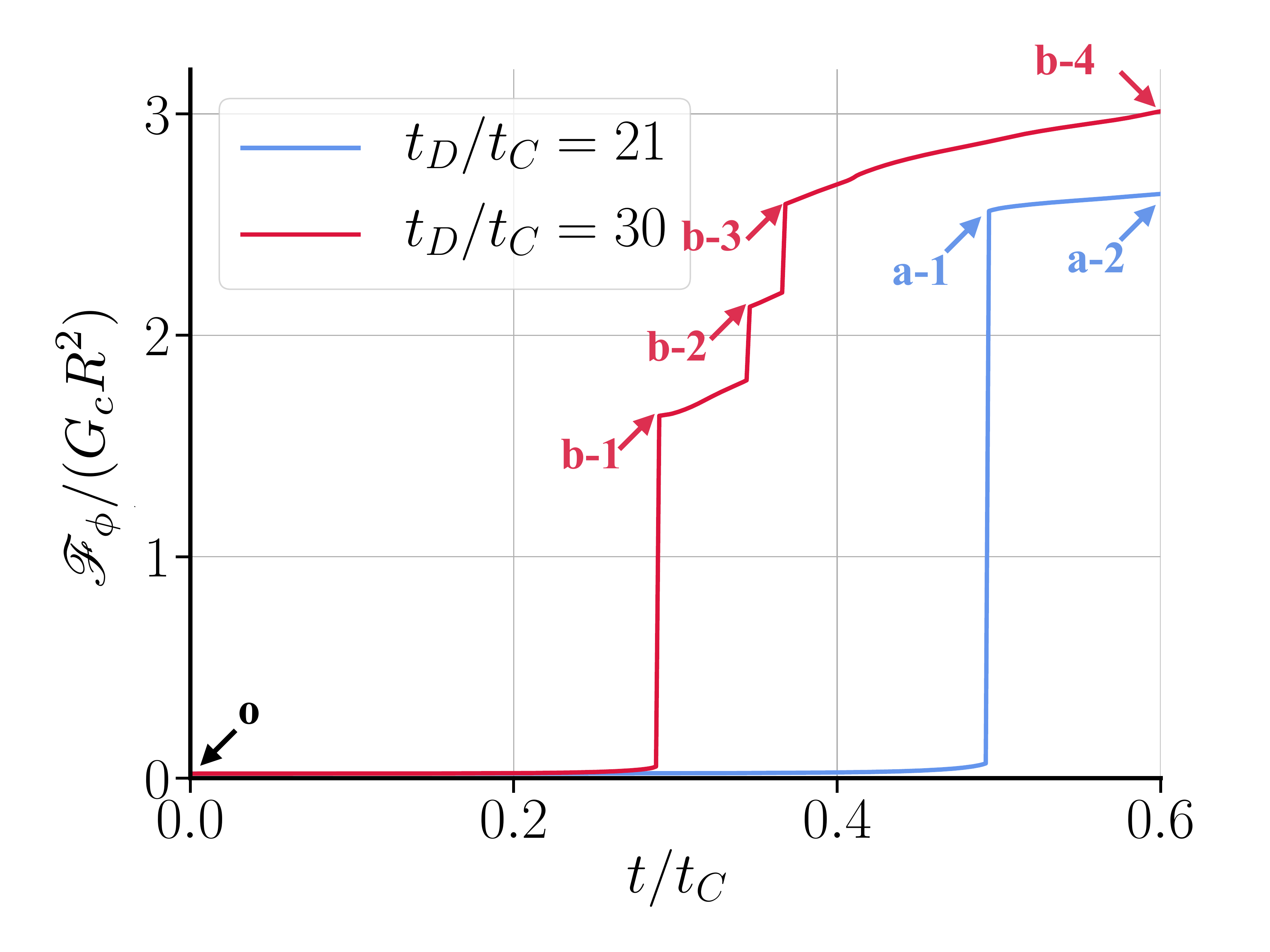}
	\end{center}
	\caption{Evolution of dimensionless surface energy $\Fs_{\phi}/(G_c\,R^2)$ for $R/l_G=2\times10^4$ spherical particle containing a $a_0/l_G=2\times10^3$ radial penny-shaped crack for $\Cr=21$ and $\Cr=30$. Associated topologies for different points in time is presented in Figs.~\protect\ref{fig:3D-R10-init-topology}--\protect\ref{fig:3D-R10-lc1-high-flux-topology}.}
	\label{fig:3D-R10-lc}
\end{figure}

\begin{figure}[htb!]
	\begin{center}
		\includegraphics[width=\columnwidth]{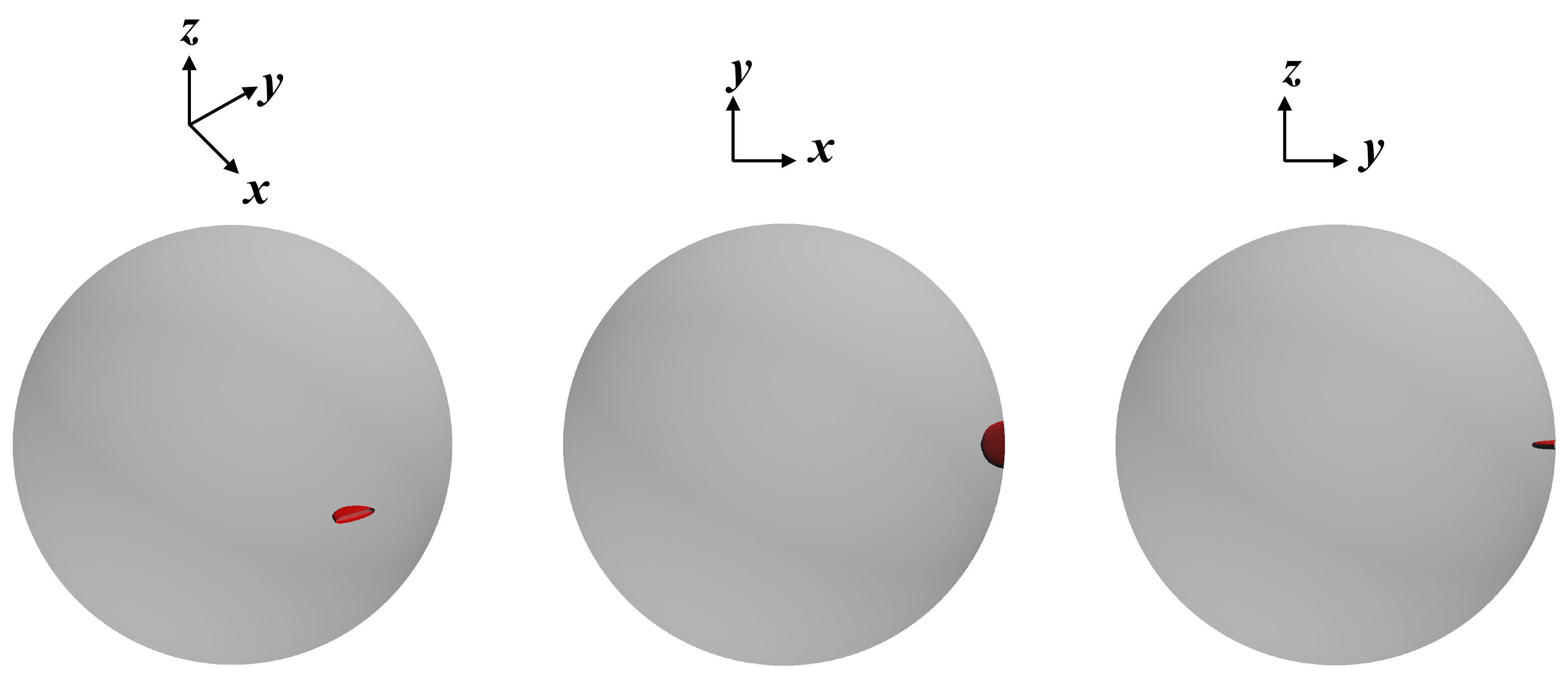}
	\end{center}
	\caption{Initial topology of the radial penny-shaped crack (iso-surface visualization for $\phi=0.5$) $a_0/l_G=2\times10^3$ in a $R/l_G=2\times10^4$ spherical particle (\textbf{o} in Fig.~\protect\ref{fig:3D-R10-lc}).}
	\label{fig:3D-R10-init-topology}
\end{figure}

\begin{figure}[htb!]
	\begin{center}
		\includegraphics[width=\columnwidth]{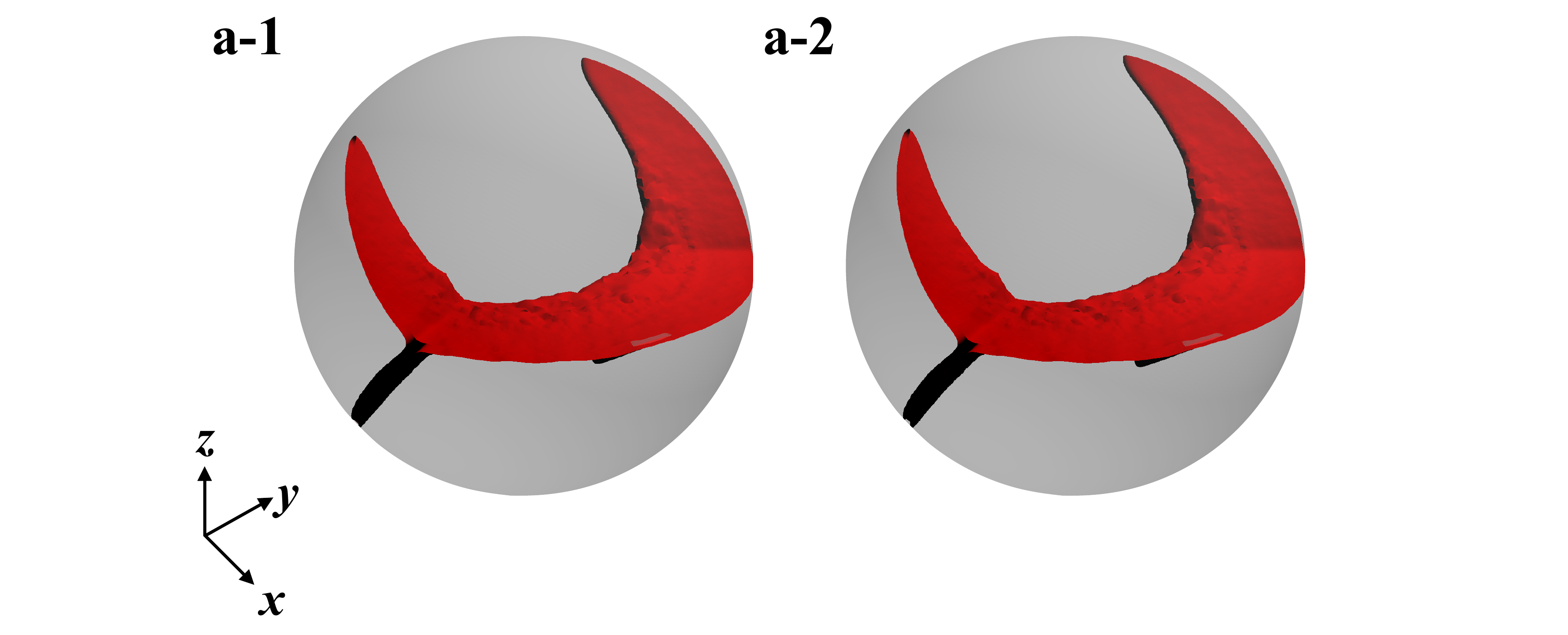}		
	\end{center}
	\caption{Evolution of high-flux fracture topology (iso-surface visualization for $\phi=0.5$) for $R/l_G=2\times10^4$ spherical particle containing an initial $a_0/l_G=2\times10^3$ radial penny-shaped crack under $\Cr=21$: initial abrupt propagation \textbf{a-1} at $t/t_C=0.5$ (left), \textbf{a-2} $t/t_C=0.6$ (right) (see Figs.~\protect\ref{fig:3D-R10-lc} and \protect\ref{fig:3D-phase-diagram}).}
	\label{fig:3D-R10-lc1-low-flux-topology}
\end{figure}

\begin{figure}[htb!]
	\begin{center}
		\includegraphics[width=\columnwidth]{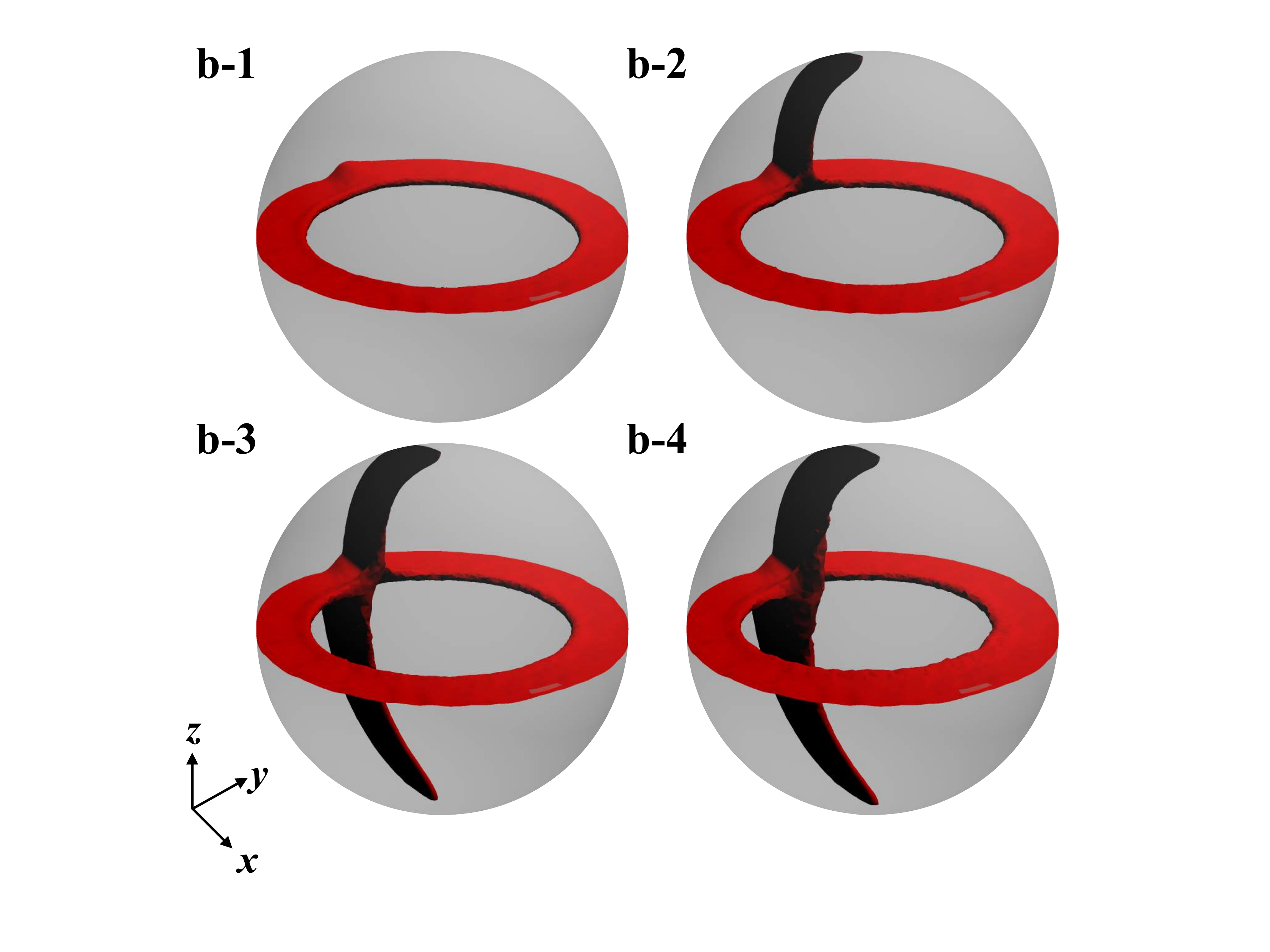}		
	\end{center}
	\caption{Evolution of low-flux fracture topology (iso-surface visualization for $\phi=0.5$) for $R/l_G=2\times10^4$ spherical particle containing an initial $a_0/l_G=2\times10^3$ radial penny-shaped crack under $\Cr=30$: initial abrupt propagation \textbf{b-1} at $t/t_C=0.28$ (top left), \textbf{b-2} $t/t_C=0.36$ (top right), \textbf{b-3} $t/t_C=0.38$ (bottom left), \textbf{b-4} $t/t_C=0.6$ (bottom right) (see Figs.~\protect\ref{fig:3D-R10-lc} and \protect\ref{fig:3D-phase-diagram}).}
	\label{fig:3D-R10-lc1-high-flux-topology}
\end{figure}

\begin{figure}[htb!]
	\begin{center}{}
		\includegraphics[width=\columnwidth]{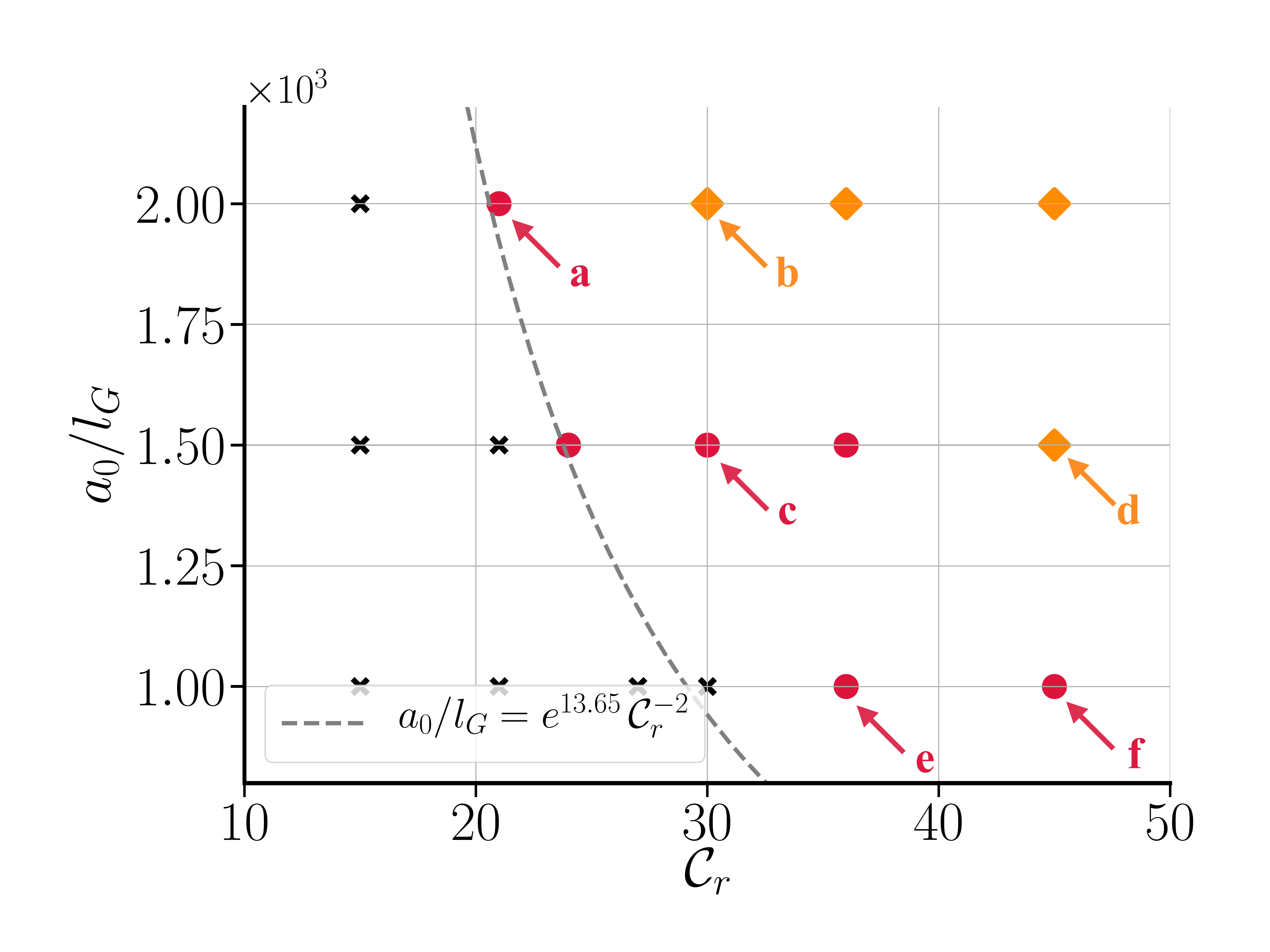}
	\end{center}
	\caption{Flaw activation diagram for a $R/l_G=2\times10^4$ radius spherical particle. Circles depict activated cracks with low-flux topology that split into multiple orientations and filled diamond depict activated cracks with high-flux topology that remain coplanar with the initial penny-shaped crack. Crosses show unactivated cracks. The gray dashed line shows the predicted power-law $a_0/l_G\sim\Cr^{-2}$ relating minimum activated flaw size and dimensionless charging rate.
}
	\label{fig:3D-phase-diagram}
\end{figure}

Fig.~\ref{fig:topology-3D} shows the complex fracture topology that results from the activation of the penny-shaped flaw in 3D.
The 3D fracture pattern highlights the role of dimensionality and follows from the fact that hoop stresses ($\sigma_{\theta\theta}$, $\sigma_{\phi\phi}$) reach their maximum values on the surface.
Consequently, the tessellation of the particle surface by the crack releases the stresses and inhibits the inward propagation of the crack.
These peripheral cracks only alleviate these stresses perpendicular to the crack surface, thereby causing new cracks to be initiated with different orientations than the plane of the initial penny-shaped crack.
Therefore, we can hypothesize that for smaller charging rates where the opening stresses (Li ions) need to penetrate farther inside, the radial propagation is augmented compared to higher charging rates where the stresses generated are more superficial.
This  hypothesis is confirmed by the results of 3D simulations presented in Fig.~\ref{fig:topology-3D}.
In that figure for all different initial flaw sizes simulated in 3D, crack propagation is abrupt and the added freedom for the cracks to release the stresses by tessellating the surface results in two dominant crack topologies: (I) cracks that propagate coplanar to the initial flaw under higher $\Cr$ (\textbf{b,d} in Figs.~\ref{fig:topology-3D}--\ref{fig:3D-phase-diagram}), and (II) cracks with initial coplanar propagation that tip split and result in a more complex topology under lower $\Cr$ (\textbf{a,c,e,f} in Figs.~\ref{fig:topology-3D}--\ref{fig:3D-phase-diagram}).

As explained before, we account for this transition using an argument similar to the one presented in Section~\ref{sub:chemo-mechanical-fracture-flux} for abrupt versus continuous propagation in 2D.
Unlike 2D radial cracks that can only release energy by penetrating towards the center of the particle, 3D penny-shaped cracks can both propagate radially and peripherally.
The radial fracture in 3D simulations is akin to the radial propagation in 2D, \ie the bulk elastic energy is released due to the crack opening up in the back.
On the other hand, the peripheral propagation is analogous to the creation of (mod) cracks in a biaxially stretched thin-films~\cite{Leon-Baldelli:2011,Leon-Baldelli:2014} or formation of imperfect polygonal patterns due to thermal quenching~\cite{Bourdin:2014a}.
Therefore, since the highest opening stresses are always created on the surface of the particle, the  initial propagation is always unstable in the peripheral direction. 
To clarify these two different fracture modes, we analyze two 3D topologies designated as cases \textbf{a} and \textbf{b} in Figs.~\ref{fig:topology-3D}--\ref{fig:3D-phase-diagram}.
Due to the complex fracture topology in 3D, we use the dimensionless surface energy $\Fs_{\phi}/(G_c\,R^2)$ as a measure of the surface area of the cracks, noticing that following equations~\eqref{eq:surface-sharp} and \eqref{eq:surface-pf}:
\begin{equation}
	\frac{\Fs_{\phi}}{G_c\,R^2}\simeq\frac{\mathcal{H}^2(\Gamma)}{R^2}
\end{equation}
Fig.~\ref{fig:3D-R10-lc} depicts the dimensionless surface energy for $a_0/l_G=2\times10^3$ at two different dimensionless charging rates: $\Cr=21$ (blue line in Fig.~\ref{fig:3D-R10-lc}) and $\Cr=30$ (red line in Fig.~\ref{fig:3D-R10-lc}).
Fig.~\ref{fig:3D-R10-init-topology} shows the initial penny-shaped crack of size $a_0/l_G=2\times10^3$ for cases \textbf{a-b}.
Similar to the arguments presented for the 2D simulations at lower $\Cr$ (cases \textbf{a,c,e,f} in Figs.~\ref{fig:topology-3D}--\ref{fig:3D-phase-diagram} and those depicted using red circles in Fig.~\ref{fig:3D-phase-diagram}), the flaw is only activated when the concentration has penetrated on the scale of the particle size.
As seen, for example, in \textbf{a-1} in Figs.~\ref{fig:3D-R10-lc} and \ref{fig:3D-R10-lc1-low-flux-topology}, the propagation of the initial flaw is first planar which then tip splits due to high biaxial stresses (due to the symmetry of the problem far from the initial flaw $\sigma_{\theta\theta}=\sigma_{\phi\phi}$).
At higher $\Cr$ (cases \textbf{b,d} in Figs.~\ref{fig:topology-3D}--\ref{fig:3D-phase-diagram} and those shown using orange diamonds in Fig.~\ref{fig:3D-phase-diagram}) the initiation is faster and creates a coplanar crack with the initial flaw as seen, for example, in \textbf{b-1} in Figs.~\ref{fig:3D-R10-lc} and \ref{fig:3D-R10-lc1-high-flux-topology}.
Upon further Li ion depletion, a secondary pair of cracks are initiated perpendicular to the initial circumferential crack as depicted in \textbf{b-2--3} in Figs.~\ref{fig:3D-R10-lc} and \ref{fig:3D-R10-lc1-high-flux-topology}.
We should highlight that the radial penetration of the 3D penny-shaped crack is similar to radial propagation in 2D simulations.
As a result, for case \textbf{a} at the lower $\Cr=21$ after the initial activation the crack abruptly penetrates radial distance of $\simeq12\times10^3\,l_G$ compared to $\simeq6\times10^3\,l_G$ for the case \textbf{b} at the higher $\Cr=30$.

Fig.~\ref{fig:3D-phase-diagram} depicts the aggregate results of a series of 3D numerical simulations for a $R/l_G=2\times10^4$ particle for $a_0/l_G=10^3,1.5\times10^3,2\times10^3$ initial penny-shaped radial flaws. 
Despite the major difference in crack propagation path (\ie penetrating cracks in two-dimensional circular particles \emph{vs.} surface cracks in the three-dimensional sphere), our 3D results suggest that critical flux to activate a surface flaw follows the inverse square power-law $a_{0,\min}\beta^2/l_G\sim\Cr^{-2}$ for moderate charging rates as in the 2D simulations.
This is not surprising since the same arguments presented in Section~\ref{sub:chemo-mechanical-fracture-flux} to justify the power-law still applies for spherical particles exposed to moderate fluxes.
Furthermore, the results in Fig.~\ref{fig:3D-phase-diagram} also suggest that, like 2D simulations (see Figs.~\ref{fig:R21-a1-Cr5-crack-propagation}--\ref{fig:R21-a1-Cr8-crack-propagation}), the transition of the inverse square power-law occurs at $a_0/l_G\simeq10^3$.

We also should note that the tiling of the sphere surface is of particular theoretical interest. 
The polygonal tiling and its number of defects is prescribed by Euler's celebrated theorem~\cite{Euler:1758}. 
In contrast, in many physical systems, the number of defects on the curved surface goes beyond the minimum number necessary and is assigned by the local energetic minima. 
Over the past decade, significant progress has been made in closely connected areas of crystal formation on spherical surfaces~\cite{Bausch:2003,Irvine:2010,Manoharan:2015} and pattern formation as the result of buckling~\cite{Jimenez:2015}. 
Although the mechanism of surface tilings generated in this section is an attractive subject for further research, in this article, we limit ourselves to the general topology of the cracks generated. 

\section{Conclusions}\label{sec:conclusion}

In this article, we developed a thermodynamically consistent framework by combining the phase-field fracture method and diffusion to model chemo-mechanical fracture.
We presented our formulation in Section~\ref{sec:formulation} and detailed our implementation of it in Section~\ref{sec:numerical-implementation}.

As our first case study, we investigated in Section~\ref{sub:chemo-mechanical-fracture-flux} the fracture of 2D circular disks.
Using different initial flaw sizes, we showed how the steep gradient created as a result of the charging rate could cause a surface flaw to propagate and fracture the particle.
Our numerical results show that for a given particle size, there exists a maximum flaw independent charging rate that can be used as a conservative limit in practice.
Furthermore, motivated by our simulation results, we showed how the activation of the surface flaws follows an inverse square law $a_{0,\min}\sim\Cr^{-2}$ over intermediate dimensionless charging rates $\Cr=O(1)$ ($t_D\sim t_C$). 
We justified this power-law behavior based on a Griffith type analysis of the stresses generated far from the crack-tip and showed how it could be used to calculate a maximum safe charging rate $C_{r,\max}$ given the elastic and fracture properties as well as an estimate of the flaw sizes in the system.
We should note that although the activation of surface flaws follows this simple power-law expression, the precise flux to activate a flaw is dictated by a non-trivial concentration profile around the crack-tip.
Since the scaling law analysis ignores the ratio of flaw size to the radius of the particle, as well as the crack-tip enrichment, the scaling constant can only be derived from the numerical simulations, especially for small particles where the initial flaw size plays a more significant role.
Furthermore, we showed that depending on the particle and flaw size, the initial propagation could be abrupt or continuous for low and high fluxes, respectively.
While puzzling at first, we described how the abrupt fracture propagation length decreases for increasing fluxes for moderate initial flaw sizes due to smaller bulk energy available at the fracture onset.

We then extended our study to high fluxes that are necessary for the activation of very small flaws ($a_0<10^3\,l_G$ for our choice of parameters).
Our results show that for these small flaws, the safe charging rate deviates from the previously obtained scaling law.
We explained our observation, noting that the high charging rate creates a depleted layer on the periphery of the particle and thus loses its effectiveness in creating a steep enough gradient to activate these flaws.
As a result, we found out that there exists a minimum safe flaw size $a_{0,\min}$ for a given particle size that does not propagate under any charging rate.
To effectively address these maximal charging rates, in Section~\ref{sub:chemo-mechanical-fracture-dirichlet}, we examined the activation of surface flaws using potentiostatic (Dirichlet) boundary conditions.
Our simulation results show that there exists a C-rate independent, safe particle size that no flaw of any size will propagate in it.
In addition, they show that in large particles the minimum activated flaw size approaches a constant value (\eg $a_{0,\min}^{\infty}\simeq200l_G$ for our choice of parameters).
In other words, our numerical simulations suggest that particles (no matter how large) containing flaws smaller than $a_{0,\min}^{\infty}$ do not crack due to diffusion-driven misfit stresses.

Finally, in Section~\ref{sub:3D-Calculations}, to investigate the role of dimensionality, we performed a series of 3D simulations on spherical particles with penny-shaped flaws.
Using our numerical observations, we showed that, unlike in 2D and assumptions~\cite{Woodford:2010} and results~\cite{Klinsmann:2016} of previous studies, the crack topology changes from a coplanar penetrating mode to a surface tiling mode.
These full (\ie without any symmetries assumed) 3D calculations show that the initial mechanical mode of failure in three-dimensional particles during charging is due to the fracture on their surface. 
While all the propagations from the initial penny-shaped crack in 3D were abrupt, we showed how the change of the fracture topology could be explained using arguments akin to those used to justify the length of abrupt propagation in 2D. 
Furthermore, our admittingly limited 3D results suggest that $a_{0,\min}\sim\Cr^{-2}$ scaling law is still valid in 3D for flaws larger than $a_0>10^3\,l_G$.

Lastly, it is crucial to highlight that, in this article, we only model chemo-mechanical fracture due to Lithium diffusion with no phase change or discontinuity in expansion. 
As highlighted, for example, in~\cite{Woodford:2012,Woodford:2013}, coherency stresses generated at the phase and grain boundaries can result in charging rate independent fracture in Li-storage materials. 

\section{Acknowledgments}
Acknowledgments: A.M. and A.K. acknowledge the support of Grant No. DE-FG02-07ER46400 from the U.S. Department of Energy, Office of Basic Energy Sciences.
The majority of the numerical simulations were performed using resources of the Extreme Science and Engineering Discovery Environment (XSEDE) under the resource allocation TG-MSS160013.
Additional numerical simulations were also performed on the Northeastern University Discovery cluster at the Massachusetts Green High Performance Computing Center (MGHPCC).

\begin{appendix}
\section{Concentration enrichment around crack-tip due to mechanical loads}\label{app:crack-tip-enrichment}
\begin{figure}[htb!]
	\begin{center}
		\includegraphics[width=\columnwidth]{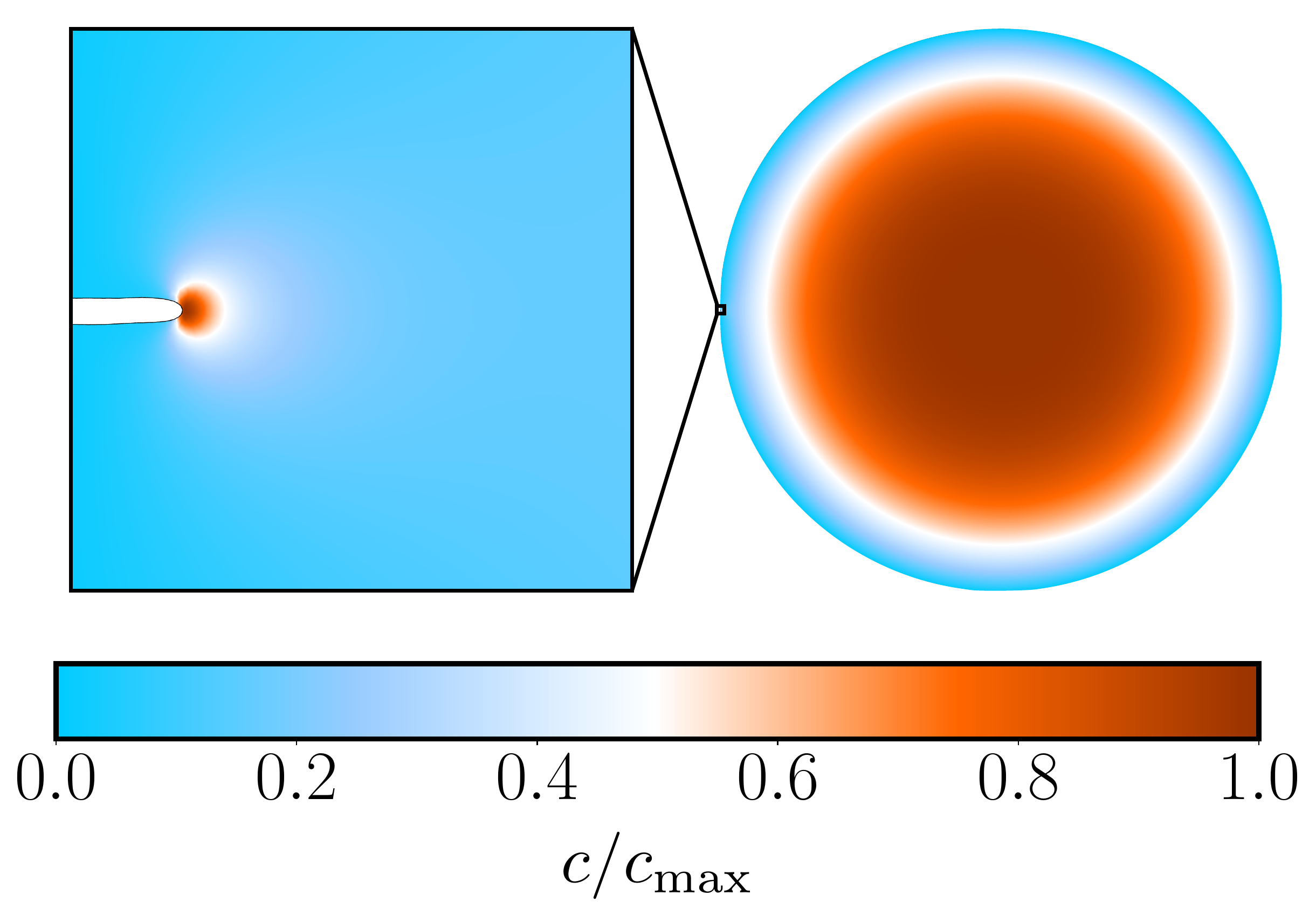}
	\end{center}
	\caption{Result of simulation for $R/l_G=4.2\times 10^4$ with a $a_0/l_G=200$ initial crack, using $\Cr=22.75$ charging rate. Color plots show the concentration at $t/t_C=1.2$:  around crack tip (left) concentration map of the particle (right).}
	\label{fig:tip-enrichment}
\end{figure}

Examining equations~\eqref{eq:diffusion}-\eqref{eq:axuilliary-diffusion}, it is easy to notice that the ions flow toward the regions with higher hydrostatic pressures; therefore, it is not surprising that in the presence of a crack, a higher concentration will accumulate at the crack-tip (see Fig.~\ref{fig:tip-enrichment} for example). 
More specifically, for small eigen-strains (\ie $\beta\ll1$) the stresses become independent of the concentration field (\ie the diffusion equation would be driven by the magnitude of hydrostatic stress). 

To derive the enrichment at the crack-tip, we can rewrite the coupled equations of elasticity and concentration in terms of Airy stress function $\mathcal{A}$ in 2-D:
\begin{align}
	\nabla^{4} \mathcal{A} &= -\El^{*} \epsilon_{0} \nabla^2 c \label{eq:airy-stress-function}
\end{align}
where $\El^{*}=\dfrac{\El}{1-\nu^2}$ for plane-stress. Therefore, for small eigen-strains \ie $\beta\ll1$ the stresses become independent of the concentration field (\ie the diffusion equation would be driven by the magnitude of hydrostatic stress). 
Thus, for steady-state conditions and in absence of a surface flux, one can write 
\begin{align}
	J&=-\nabla\frac{\delta F}{\delta c}= 0 \nonumber \\
    \frac{\delta F}{\delta \bar{c}}&=\cm\R T\ln\left(\frac{c}{\cm-c}\right)-\cm\epsilon_{0}\Tr(\sigma)=\mu_{0} \\
    c&=\frac{\cm}{1+\exp\left(-\dfrac{\beta {\sigma}_{kk}+{\mu}_{0}}{\cm \R T}\right)}\label{eq:solution-c-Fermi}
\end{align}

A similar solution to~\eqref{eq:solution-c-Fermi} can be obtained for the dilute approximation where $f(c)=c\ln(c)-c$ as:
\begin{equation}
	{c}=\cm \exp\left(\dfrac{\beta {\sigma}_{kk}+{\mu}_{0}}{\cm \R T}\right)\label{eq:solution-c-exp}
\end{equation}
The above expression was also derived by the direct solution of the diffusion equation in~\cite{Liu:1970}.

To find the concentration profile around the crack-tip, we can replace the expression of $\Tr(\sigma)$ from the asymptotic plane-stress solution of mode-I fracture:
\begin{align}
\sigma_{xx}(r,\theta)&=\frac{K_I}{\sqrt{2\pi r}}\cos\left(\frac{\theta}{2}\right)\left[1-\sin\left(\frac{\theta}{2}\right)\sin\left(\frac{3\theta}{2}\right)\right]+O(\sqrt{r})\label{eq:crack-tip-sxx}
\end{align}
\begin{align}
\sigma_{yy}(r,\theta)&=\frac{K_I}{\sqrt{2\pi r}}\cos\left(\frac{\theta}{2}\right)\left[1+\sin\left(\frac{\theta}{2}\right)\sin\left(\frac{3\theta}{2}\right)\right]+O(\sqrt{r})\label{eq:crack-tip-syy}
\end{align}
where $K_I$ is the stress intensity factor. After some algebra $\Tr(\sigma)$ can be written as:
\begin{equation}
	\Tr(\sigma(r,\theta))=\frac{2 K_I}{\sqrt{2\pi r}} \cos\left(\frac{\theta}{2}\right)\label{eq:Tr-sigma-mode-I}
\end{equation}
where $K_{I}$ is the mode-I stress intensity factor. Using~\eqref{eq:Tr-sigma-mode-I} we can write the concentration around the crack-tip at the time of fracture as:
\begin{equation}
	c=\frac{\cm}{1+exp\left(-\sqrt{\dfrac{2r_c}{\pi r}}\left(\dfrac{K_I}{K_{IC}}\right)\cos\left(\dfrac{\theta}{2}\right)-\bar{\mu}_{0}\right)}\label{eq:solution-c-Fermi-Mode-I}
\end{equation}
where 
\begin{equation}\label{eq:rc}
	r_c=\left(\frac{\beta K_{IC}}{\cm \R T}\right)^2=l_G\left(\frac{\beta\El}{\cm\R T}\right)^2
\end{equation}
can be identified as the intrinsic length scale for the concentration of ions around the crack-tip. 
Equation~\eqref{eq:rc} shows that the ratio of the enrichment length scale to Griffith length scale scales as the square ratio of maximum misfit stresses to chemical energy.
In~\eqref{eq:solution-c-Fermi-Mode-I} one can find the steady-state chemical potential $\mu_{0}$, from far field concentration as
\begin{equation}
 	{\mu}_{0}=\cm \R T\ln\left({{c}_{\infty}}/{(\cm-{c}_{\infty})}\right)
 \end{equation} 

\begin{figure}[htb!]
	\begin{center}
		\includegraphics[width=\columnwidth]{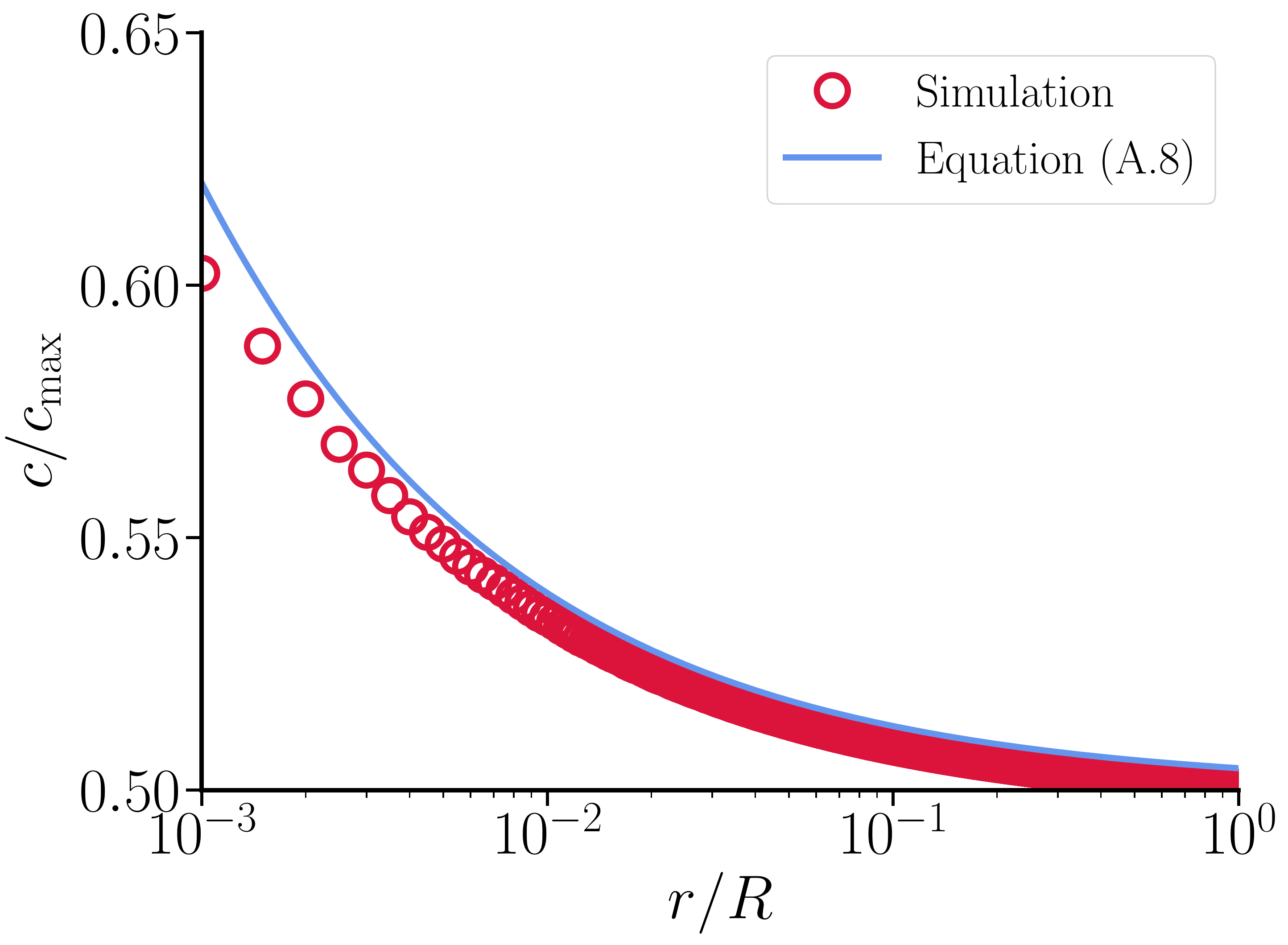}
	\end{center}
	\caption{Comparison of relative concentration $c/\cm$ near crack-tip for $\beta=0.025$: numerical simulation using 2D simulations of circular geometry with asymptotic near crack-tip displacement boundary conditions (red circles), closed-form solution~\protect\eqref{eq:solution-c-Fermi-Mode-I}.}
	\label{fig:comparison-rc}
\end{figure}

As we showed in the Section~\ref{sub:chemo-mechanical-fracture-flux}, crack-tip enrichment is a common occurrence in diffusion-driven fracture of Lithium-ion battery particles. 
We can easily calculate the length scale $r_c/l_G\simeq7591.75$ for \ce{LiMn2O4} at room temperature where the crack-tip concentration is captured approximately by~\eqref{eq:solution-c-Fermi-Mode-I}. 
Fig.~\ref{fig:comparison-rc} shows a comparison between the results of the numerical simulation for a $R/l_G=4.2\times 10^4$ particle (Fig.~\ref{fig:R21-a5-crack-propagation}) and~\eqref{eq:solution-c-Fermi-Mode-I}.
The simulation is performed in a circular geometry of radius $R$ containing a sharp $1^{\circ}$ notch from $r=-150\,\xi$ to $r=0$ at $\theta=\pi$.
To simulate near tip stress fields, the displacement fields associated with~\eqref{eq:crack-tip-sxx}--\eqref{eq:crack-tip-syy} were imposed on the boundary of the domain.
The concentration is initially uniform $c/\cm=0.5$ everywhere and the value of $\mu_0$ was calculated based on the resulting concentration at $t/t_D=1$ and $r=R$. 

We should note that the radial crack, driven by charging the cathodic particle, can stop propagating in the middle of the particle. In this situation the enrichment carried by the crack-tip can be shielded from the depleting flux by chemo-mechanical force exerted at the crack-tip. 
The remaining concentration then can change the dynamics of the charging process. 
Furthermore, while the main focus of this article is on the diffusion of Li-ions in battery particles, crack-tip enrichment can play an important role in other systems where diffusion and fracture happen concurrently such as corrosive cracks, crack-tip embrittlement, and fracture in poroelastic media~\cite{Song:2013,Bouklas:2015}. 

\end{appendix}
\bibliographystyle{unsrt}
\bibliography{Mesgarnejad-Karma-Li-ion1}

\end{document}